\documentclass[a4paper]{article}

\usepackage{amsfonts}

\usepackage{slashed}

\def\be{\begin{equation}}
\def\ee{\end{equation}}
\def\bea{\begin{eqnarray}}
\def\eea{\end{eqnarray}}
\def\({\left(}
\def\){\right)}
\def\<{\left<}
\def\>{\right>}

\def\>{\rangle}
\def\<{\langle}
\def\|{\mid}

\def\tr{{\mbox{tr}}}
\def\be{\begin{equation}}
\def\ee{\end{equation}}
\def\bea{\begin{eqnarray*}}
\def\eea{\end{eqnarray*}}
\def\ben{\begin{eqnarray}}
\def\een{\end{eqnarray}}
\def\({\left(}
\def\){\right)}
\def\<{\left<}
\def\>{\right>}

\def\[{\left[}
\def\]{\right]}

\def\+{\bar}
\def\mb{\mathbb}
\def\tr{{\mbox{tr}}}

\def\L{{\cal{L}}}

\def\t{\tilde}

\def\t{\widetilde}

\def\n{{\cal{N}}}
\def\O{{\cal{O}}}

\def\E{{\cal{E}}}

\def\ee{\breve{e}}

\def\+{\breve{+}}
\def\-{\breve{-}}

\begin{document}
\setlength{\unitlength}{1mm}

\pagestyle{empty}
\vskip-10pt
\vskip-10pt
\hfill 
\begin{center}
\vskip 3truecm
{\Large \bf
M5 brane on ${\mathbb{R}}^{1,2} \times S^3$}\\
\vskip 2truecm
{\large \bf
Andreas Gustavsson\footnote{a.r.gustavsson@swipnet.se}}\\
\vskip 1truecm
{\it  Physics Department, University of Seoul, 13 Siripdae, Seoul 130-743 Korea}
\end{center}
\vskip 2truecm
{\abstract{We deconstruct $16$ rigid supersymmetry variations for M5 brane on $\mb{R}^{1,2}\times (S^3/{\mb{Z}_k})$ and obtain on-shell closure on Lie derivatives. Dimensional reduction on the Hopf fiber by taking $k\rightarrow \infty$ gives sYM on ${\mb{R}}^{1,2} \times S^2$ with $8$ rigid supersymmetries. We reproduce the result in arXiv:0908.3263 but we also derive an additional graviphoton term which is associated with the twisting of the Hopf bundle.}}

\vfill 
\vskip4pt
\eject
\pagestyle{plain}

\section{Introduction}
A deeper insight in a theory may be obtained by studying some deformation of the theory. Some features of the theory may be visible only in the deformed theory and some regularization techniques may require that we deform the theory. In this paper we will study M5 brane\footnote{By M5 brane theory we refer to the six-dimensional up-lift of ${\cal N} = 4$ sYM theory in four dimensions, to which it reduces upon dimensional reduction.} on ${\mb{R}}^{1,2} \times S^3$ which can be deconstructed from a mass deformation of BLG theory. The mass parameter $m$ of deformed BLG theory is related to the radius $R$ of $S^3$ as 
\bea
R &=& -\frac{1}{m}
\eea
and the $S^3$ is generated by a Meyers effect \cite{Myers:1999ps}. For the M5 brane we will see that it can be made maximally supersymmetric when including certain correction terms of order $\frac{1}{R}$ and $\frac{1}{R^2}$. This is a bit surprising since the partially curved six-manifold ${\mb{R}}^{1,2} \times S^3$ can not be conformally mapped into $\mb{R}^{1,5}$. We expect the M5 brane theory on $\mb{R}^{1,2} \times S^3$ is unique and it would be interesting to see if the theory can be derived using the method in \cite{Festuccia:2011ws} though we will not attempt this here.

In \cite{Nastase:2009zu} the D4 brane is deconstructed on ${\mb{R}}^{1,2} \times S^2$ from matrix realization of mass deformed ABJM theory by taking the large $N$ and the large $k$ limit. In this paper we will recover this D4 brane theory by dimensionally reducing the M5 brane theory on ${\mb{R}}^{1,2} \times (S^3/{\mb{Z}_k})$ by taking $k\rightarrow \infty$ which effectively shrinks the length of the $U(1)$ Hopf fiber by a factor of $\frac{1}{k}$. The dimensional reduction along the Hopf fiber breaks half of the supersymmetry. 

It appears that the deconstruction used in \cite{Nastase:2009zu} does not give the complete answer. It misses out the graviphoton term 
\bea
-\frac{k}{8\pi^2} \int V \wedge F \wedge F
\eea
in the D4 brane action. Here $V$ is the graviphoton, which in our situation is the non-trivial connection one-form of the $U(1)$ bundle over $S^2$ which makes it correspond to $S^3$, and $F = dA$ is the Maxwell field strength. We will show that the graviphoton term arises by dimensional reduction of M5 brane on the Hopf fiber. We will also show that this term is necessary in order for the D4 brane action to be supersymmetric.\footnote{I would like to thank Martin Cederwall for asking the question if supersymmetry can be used to derive this term. However, this term can not be deduced by supersymmetry for cases when $V$ is a flat connection since this term would be supersymmetric by itself. It is then a topological term which is invariant under any deformations that are continuously connected to the identity map. Locally it is a total derivative, but not globally since the harmonic part of $F$ is not globally expressible as $dA$. The harmonic part of $F$ is invariant under a supersymmetry variation, which can be understood from the fact that $F$ does not have a conjugate momentum variable (this was pointed out to me by M\r{a}ns Henningson), hence taking the Poisson bracket of $F$ and any conserved charged of the theory necessarily gives a vanishing variation of $F$.}

The results of this paper can be compactly summarized as we do in section \ref{summary}. However, we think that the methods we use to derive these results (deconstruction and dimensional reduction respectively) are interesting on their own.

A previous work \cite{Gustavsson:2009qd} also deals with M5 brane on ${\mb{R}}^{1,2}\times S^3$. This work contains some crucial numerical errors and the explicit check of supersymmetry is missing. The expression for the Nambu bracket was not rigorously derived when one of its entries is a spinor. In this paper we remedy these issues.

\section{Mass deformed BLG theory}
There is a mass deformation of original BLG theory which preserves all $16$ supersymmetries \cite{Hosomichi:2008qk}, \cite{Gomis:2008cv}. If we denote a yet unspecified gauge invariant inner product by the bracket $\<\bullet,\bullet\>$,  by $T^a$ some real yet unspecified three-algebra generators, and the three-bracket $[\bullet,\bullet,\bullet]$, which is totally antisymmetric and satisfies the fundamental identity, then the undeformed Lagrangian with manifest $SO(8)$ symmetry and $\n=8$ supersymmetry, is given by \cite{Bagger:2007jr}
\bea
\L &=& -\frac{1}{2}\<D_{\mu}X^I,D^{\mu}X^I\> - \frac{1}{12} \<[X^I,X^J,X^K],[X^I,X^J,X^K]\>\cr
&&+ \frac{i}{2}\<\bar{\psi},\Gamma^{\mu}D_{\mu}\psi\> + \frac{i}{4}\<\bar{\psi},\Gamma_{IJ}[\psi,X^I,X^J]\>\cr
&&-\frac{1}{2}\epsilon^{\mu\nu\lambda}\<T^a,[T^b,T^c,T^d]\}\> A_{\mu,ab} \partial_{\nu} A_{\lambda,cd}
\eea
The mass deformation is given by
\bea
\L_m &=& -\frac{m^2}{2}\<X^I,X^I\> + \frac{im}{2} \<\bar{\psi},\Gamma_{(4)} \psi\>\cr
&& + \frac{m}{6} \(\epsilon_{ijkl}\<X^i,[X^j,X^k,X^l]\> + \epsilon_{\hat{i}\hat{j}\hat{k}\hat{l}} \<X^{\hat{i}},[X^{\hat{j}},X^{\hat{k}},X^{\hat{l}}]\>\)
\eea
The deformed Action is invariant under deformed $\n=8$ supersymmetry, but only has $SO(4) \times SO(4)$ R-symmetry. Accordingly we split the $SO(8)$ vector index as $I = (i,\hat{i})$, and we define $\Gamma_{(4)} = \Gamma_{\hat{1}\hat{2}\hat{3}\hat{4}}$. The deformed supersymmetry variations are given by
\bea
\delta X^I &=& i\bar{\epsilon} \Gamma^I \psi\cr
\delta \psi &=& \Gamma^{\mu}\Gamma_I \epsilon D_{\mu} X^I - \frac{1}{6} \Gamma_{IJK} \epsilon [X^I,X^J,X^K] - m \Gamma_{(4)} \Gamma_I \epsilon X^I\cr
\delta A_{\mu} &=& i\bar{\epsilon} \Gamma_{\mu}\Gamma_I [\bullet,X^I,\psi]
\eea
Infinitesimal gauge variations act on the fields according to
\bea
\delta X^I &=& \Lambda_{ab}[T^a,T^b,X^I]\cr
\delta \psi &=& \Lambda_{ab}[T^a,T^b,\psi]\cr
\delta A_{\mu,ab} &=& D_{\mu} \Lambda_{ab}
\eea
The covariant derivative is given by
\bea
D_{\mu} X^I &=& \partial_{\mu} X^I - [X^I,T^c,T^d] A_{\mu,cd}
\eea
The supersymmetry parameter and the fermion in the theory have opposite chiralities
\bea
\t \Gamma \epsilon &=& \epsilon\cr
\t \Gamma \psi &=& -\psi
\eea
Here
\bea
\t \Gamma &=& \Gamma_{012}
\eea
and we use eleven-dimensional gamma matrices.

So far we have not restricted ourself to any particular three-algebra. There are infinitely many infinite-dimensional metric three-algebras with a positive-definite metric $\<T^a,T^b\>$ that we may consider. Namely to any three-manifold $\n$ we can associate a three-bracket defined as the Nambu bracket on $\n$,
\bea
[T^a,T^b,T^c] &=& \{T^a,T^b,T^c\}
\eea
where
\bea
\{T^a,T^b,T^c\} &=& \omega^{\alpha\beta\gamma} \partial_{\alpha} T^a \partial_{\beta} T^b \partial_{\gamma} T^c
\eea
is the Nambu bracket, and where 
\bea
\omega_{\alpha\beta\gamma} &=& \sqrt{g}\epsilon_{\alpha\beta\gamma}
\eea
is the totally antisymmetric tensor, constructed out of the Levi-Civita symbol $\epsilon_{\alpha\beta\gamma} = \pm 1$ and the determinant of the metric tensor on $\n$, $g = \det g_{\alpha\beta}$, which are both tensor densities. The above combination is such that the weights of these tensor densities cancel, leaving us with a totally antisymmetric tensor which we denote by $\omega_{\alpha\beta\gamma}$. Being a tensor, we can define $\omega^{\alpha\beta\gamma}$ again as a tensor by rising its indices by means of the metric tensor. We take as three-algebra generators any complete set of real-valued functions on $\n$. The positive-definite inner product is given by
\bea
\<T^a,T^b\> &\sim & \int d^3 \sigma \sqrt{g} T^a T^b
\eea
All requirements of a real three-algebra are satisfied by this choice of three-bracket and inner product for any choice of $\n$, which means that we can try to associate a BLG theory to any choice of $\n$. Let us refer to such a theory as a Nambu-BLG($\n$) theory \cite{arXiv:0805.2898}. 

We will now proceed to check whether supersymmetry could put restrictions on $\n$. First of all we must assure that
\bea
\{\bar{\epsilon}\Gamma^I \psi, X^J,X^J\} &= & \bar{\epsilon} \Gamma^I \{\psi,X^J,X^K\}
\eea
This is a trivial identity if we work with a matrix realization of the BLG theory three-algebra (and $SO(4)$ is of course the only relevant example which has a positive definite metric). For functions on $\n$ this identity may at first sight not seem to be that obvious. However it is obvious, because these spinor entities $\epsilon$ and $\psi$ are defined on flat euclidean transerse space (and let us suppress the space-time dependence in our discussion here), and in particular the supersymmetry parameter in BLG theory does not depend on the coordinates $x^I$ of the transverse space 
\bea
\partial_I \epsilon &=& 0
\eea
Our viewpoint is that $X^I$ describes the position of the M2 branes in the transverse space. Using that constant spinor, we can project onto a derivative along $\n$. If $\n$ is parametrized as follows
\bea
\sigma^{\alpha} &\mapsto & T^I(\sigma)
\eea
then we may define a tangential derivative to $\n$ as
\bea
\partial_{\alpha} \epsilon &:=& \frac{\partial T^I}{\partial \sigma^{\alpha}} \partial_I \epsilon
\eea
and then this tangential derivative will also vanish. The Nambu bracket with an euclidean spinor entity $\psi$ is defined just as usual, thus for instance
\bea
\{\psi,X^J,X^K\} &=& \omega^{\alpha\beta\gamma} \partial_{\alpha}\psi \partial_{\beta}X^J \partial_{\gamma} X^K
\eea
It may be noted that the dependence on $\sigma^{\alpha}$ of the spinor $\psi$ comes entirely from the three-algebra generators, as $\psi_a(x^{\mu})T^a(\sigma^{\alpha})$, and that is part of the deconstruction idea. We get a six-dimensional quantity from a three-dimensional one. The BLG supersymmetry parameter is not a spinor that lives on the submanifold $\n$ but it rather lives on flat euclidean transverse space ${\mb{R}}^8$. To get from $\epsilon$ to a spinor on $\n$ we perform a further transformation which is explained in detail in the appendices of this paper. This further transformation involves a certain transition matrix $g$ (which can be constructed out of a vielbein on $\n$ and the tangential and normal derivatives to $\n$, that is $\partial_{\alpha}T^I$ and $\partial_A T^I$ where $A$ labels five normal coordinates), and we can then define the spinor $\t \epsilon$ which lives on $\n$ according to
\bea
\epsilon &=& g \t \epsilon
\eea
From the condition that $\epsilon$ is constant, we then get
\ben
\(\partial_{\alpha} + g^{-1}\partial_{\alpha} g \)\t\epsilon &=& 0\label{cons}
\een
Now we want to express this condition in terms of a covariant derivative on $\n$. The covariant derivative on $\n$ involves the spin connection $\Omega_{\alpha}$ on $\n$, and if $\n$ is curved the spin connection is not flat. So clearly $\Omega_{\alpha} \neq g^{-1}\partial_{\alpha} g$ in general. The question now arises, when can we express Eq (\ref{cons}) in terms of the covariant derivative on $\n$? We can rewrite (\ref{cons}) as follows,
\bea
D_{\alpha}\t\epsilon &=& \(\Omega_{\alpha} - g^{-1}\partial_{\alpha} g\)\t\epsilon
\eea
We note that once we choose $\n$ everything including the transition matrix $g$ and the spin connection $\Omega_{\alpha}$ are uniquely determined, so there is no further freedom to adjust anything in this equation once $\n$ has been chosen. Since covariant derivatives do not commute, this set of equations now provide us with a non-trivial integrability condition. For most choices of $\n$ these equations can not be integrated to give us a solution $\t\epsilon$. Since the BLG spinor has $8$ spinor components along the transverse space, and only $2$ of them can be associated to $\n$, we have to require the existence of $4$ Killing spinors on $\n$. Another, and possibly very interesting, class of solutions might be provided by the conformal Killings spinors \cite{Linander:2011jy}, although we will not consider this possibility any further in this paper. 

The existence of four independent Killing spinors $\epsilon^X$ ($X=1,...,4$) implies the existence of six associated independent Killing vectors
\bea
V^{XY}_{\alpha} &=& \bar{\epsilon}^X \Gamma_{\alpha} \epsilon^Y
\eea
since the right-hand side is antisymmetric under exchange of $X$ and $Y$. We have thus found that $\n$ must have six independent Killing vectors, which means that $\n$ must be maximally symmetric. Up to discrete identifications, $\n$ can then only be either flat euclidean space, a three-sphere or a de Sitter space. 

So far we have only studied restrictions on $\n$ that we get by requiring supersymmetry of the theory. In order to be able to deconstruct a maximally supersymmetric M5 brane theory from Nambu-BLG($\n$) we must in addition require that $\n$ preserves all the supersymmetries of the Nambu-BLG($\n$) theory. There are at least two different ways to achieve this. One way is by utilizing a space-time independent shift symmetry of the spinor in BLG theory \cite{arXiv:0805.2898}, \cite{arXiv:0806.4044}, \cite{arXiv:1006.5291}, \cite{arXiv:1008.0902}, \cite{arXiv:1012.2707}. The other way is by mass deforming the BLG theory in a way that preserves all the supersymmetry. In this paper we will focus on the latter alternative. 

As far as we can understand, it is not possible to formulate a Nambu-BLG($\n$) theory on an arbitrary three-manifold $\n$. In order to extend this to general $\n$, one may instead consider twisted (or partially twisted) BLG theory and then one may be able to consistently define the three-bracket as a Nambu bracket on such an $\n$. From this one may also attempt to also deconstruct twisted (or partially twisted) M5 brane theory. In the maximally supersymmetric case with $\n=S^3$, twisting is not mandatory, but of course it is possible to twist also for this case, though that would not give us anything new but it would be just a reformulation.

\subsection{Supersymmetric three-sphere vacuum solution} 
A static and maximally supersymmetric vacuum solution in massive BLG theory can be found by solving 
\bea
\delta \psi &=& 0
\eea
by taking $X^I = T^I$ where $T^{\hat{i}} = 0$ and 
\ben
[T^i,T^j,T^k] &=& -m \epsilon^{ijkl} T^l\label{vacuum}
\een
It is easy to check that whenever Eq (\ref{vacuum}) is satisfied, such a field configuration preserves all the $\n=8$ mass-deformed supersymmetries. One can also check that the Action vanishes on this solution. Vanishing Action of course is the same as vanishing Hamiltonian since the solution is static. 

We know of only two ways to solve Eq (\ref{vacuum}). The first way is to take the gauge group to be $SO(4)$. The second way is to take the gauge group to be the the infinite-dimensional group of volume preserving diffeormorpisms (VPD) of a round $S^3$ embedded in flat euclidean $\mb{R}^4$, which in turn is embedded into transverse space as ${\mb{R}}^8 = {\mb{R}}^4 \times {\mb{R}}^4$, where we choose the convention such that we embed $S^3$ into the first of these two ${\mb{R}}^4$ factors. There is nothing in between these two ultimate cases, and in particular one can not reach the VPD's by a limiting procedure from finite-rank gauge groups. But this does not have to mean that the realization in terms VPD's is less interesting or its study would have to be less rigorous than realizations in terms of finite-rank matrices of finite-rank gauge groups. One aim of this paper is precisely that, to show that very precise results (such as the precise value of the M5 brane coupling constant, as well as the precise form of its action) can be obtained by starting with BLG theory realized by a VPD gauge group on $S^3$.

In order to understand how we can realize Eq (\ref{vacuum}) by taking the gauge group of VPD's on $S^3$, we first notice that if $x^i$ denote the Cartesian coordinates in ${\mb{R}}^4$, then
\bea
*_4(dx^i \wedge dx^j \wedge dx^k) &=& \epsilon^{ijkl} dx^l
\eea
Let us now consider a coordinate transformation from Cartesian to Spherical coordinates $x^i = T^i(\sigma^{\alpha},R)$ with metric
\bea
dx^i dx^i &=& g_{\alpha\beta} d\sigma^{\alpha} d\sigma^{\beta} + (dR)^2
\eea
Let us then expand the differential in tangential and normal components,
\bea
dx^i &=& d\sigma^{\alpha} \partial_{\alpha} T^i + dR \frac{T^i}{R}
\eea
and plug back it back into the duality relation, and separate out the $dR$ components using that 
\bea
*_4(d\sigma^{\alpha} \wedge d\sigma^{\beta} \wedge d\sigma^{\gamma}) &=& \omega^{\alpha\beta\gamma} dR
\eea
We then get
\ben
\{T^i,T^j,T^k\} &=& \frac{1}{R} \epsilon^{ijkl} T^l\label{sp}
\een
We now see that by making the identification
\ben
R &=& - \frac{1}{m}\label{m}
\een
we get a fairly concrete realization of the vacuum equation (\ref{vacuum}), even though it is given in terms of functions $T^i = T^i(\sigma^{\alpha})$ rather than in terms of some matrices $T^i$. \footnote{One may wonder what happens if instead of $dR$ one would pick another component, which is $d\sigma^{\alpha}$. If we do that, we will not produce the Nambu bracket, but rather we get 
\ben
\epsilon^{ijkl}\omega^{\alpha\beta}{}_{\gamma} \partial_{\alpha}x^i \partial_{\beta}x^j x^k &=& -2R \partial_{\gamma} x^l\label{ett}
\een
This is not an independent equation that shall be also satisfied for any arbitrary parametrization of the $S^3$ embedded into ${\mb{R}}^4$. This relation can instead be derived from the relation 
\ben
\epsilon^{ijkl} \partial_{\alpha} x^i \partial_{\beta} x^j \partial_{\gamma}x^k &=& \frac{1}{R}\omega_{\alpha\beta\gamma} x^l\label{tva}
\een
which in turn follows from the relation Eq (\ref{sp}). We then derive Eq (\ref{tva}) by contracting Eq (\ref{tva}) by $x^l \partial_{\delta} x^k \omega^{\alpha\beta\tau}$ and by using $x^l x^l = R^2$. In short, Eq (\ref{ett}) is derived from the Nambu bracket relation and the sphere constraint.}

In BLG($S^3$) theory we can realize the 3-algebra generators $T^a$ explicitly by the spherical harmonics
\bea
c_{i_1...i_n} T^{i_1}...T^{i_n}.
\eea
Here the independent ones are obtained from traceless symmetric coefficients $c_{i_1...i_n}$, and we must take $n=1,2,...,\infty$ since there is no consistent finite-dimensional trunction of the associated three-algebra. We define the 3-bracket for any three 3-algebra generators $T^a$, $T^b$ and $T^c$ in terms of the Nambu bracket defined on $S^3$,

To see that gauge variations correspond to VPS's, let us recall that a gauge transformation acts on a matter field $X(\sigma^{\alpha}) = X_a T^a(\sigma^{\alpha})$ as
\bea
\delta X(\sigma^{\alpha}) &=& \{T^a,T^b,T^c\}(\sigma^{\alpha}) X_c
\eea
which we may write in the form
\bea
\delta X(\sigma^{\alpha}) &=& v^{\alpha} \partial_{\alpha} X(\sigma^{\alpha})
\eea
where
\bea
v^{\alpha} &=& \omega^{\alpha\beta\gamma} \partial_{\beta}T^a \partial_{\gamma} T^b
\eea
For this parameter we can compute
\bea
D_{\alpha} v^{\alpha} &=& \omega^{\alpha\beta\gamma} \(D_{\alpha}\partial_{\beta} T^a \partial_{\gamma}T^b + \partial_{\beta} T^a D_{\alpha}\partial_{\beta} T^b\)
\eea
and this vanishes because 
\bea
D_{\alpha}\partial_{\beta} T^i &=& D_{\beta} \partial_{\alpha} T^i
\eea
and by the antisymmetry of $\omega^{\alpha\beta\gamma}$. It means that we consider diffeomorphisms 
\bea
\sigma^{\alpha} &\mapsto & \sigma^{\alpha} - v^{\alpha}
\eea
such that 
\bea
D_{\alpha} v^{\alpha} & =& 0
\eea
These are precisely those diffeomorphisms that leave the determinant of the metric tensor invariant.

Later we will also derive the explicit expression for the Nambu bracket when one of its entries is a spinor on $\n = S^3$. To this end we will use the vielbein formalism developed in \cite{Abrikosov:2002jr} to transform a spinor from Cartesian to Polar coordinates on ${\mb{R}}^4 \simeq S^3 \times \mb{R}_+$. The same spinor formalism was used in \cite{Nastase:2009zu} to deconstruct D4 on $S^2$, by mimicing the previous work \cite{Andrews:2006aw} where twisted Maldacena-Nunez compaction of $(1,1)$ gauge theory on ${\mb{R}}^{1,3} \times S^2$, was deconstructed from $N=1^*$ sYM. Using this spinor formalism, we deconstruct the untwisted M5 brane theory from massive BLG theory. We are then able to check closure of the untwisted supersymmetry variations. We find that they are of a standard form, and closure nicely comes out with all its Lie derivatives on $S^3$.

Ideally one would like to have a direct derivation of the emergence of the Nambu bracket on $S^3$ as the large $N$ limit of some discrete version of the Nambu bracket. We believe that ABJM theory with gauge group $U(N) \times U(N)$ \footnote{or maybe $U(N)\times U(N-1)$} provides the discrete version of our BLG theory with Nambu bracket. If this is correct, it should have the profound implication that D4 brane that can be deconstructed out of the large $N$ limit of ABJM theory \cite{Nastase:2009zu}, should be dual to M5 brane which we construct out of BLG theory with a Nambu bracket in this paper. This also shows how difficult it has to be to understand the emergence of the Nambu bracket and the BLG theory as the large $N$ limit. It is presumably just as difficult as showing the duality between D4 brane and M5 brane \cite{Lambert:2010iw}, \cite{Douglas:2010iu}.

\section{Deconstructing M5 brane}

\subsection{Fluctuations}
We expand the scalars fields around the three-sphere vacuum in tangential and transverse fluctuation fields
\bea
X^I &=& T^I + Y^{\alpha} \partial_{\alpha} T^I + \frac{Y}{R} T^I + Y^{\hat{i}} \partial_{\hat{i}} T^I
\eea
We may dualize the tangential components into a two-form and we may use the gauge potential to deconstruct $\mu\alpha$ components of this two-form. Thus we define
\bea
B_{\alpha\beta} &=& \omega_{\alpha\beta\gamma} Y^{\gamma}\cr
B_{\mu\alpha} &=& A_{\mu,ab} T^a \partial_{\alpha} T^b
\eea
In this fluctuation expansion we have evaluated the radius derivative as $\partial_R T^I = \frac{1}{R} T^I$.

The BLG spinors $\psi$ and $\epsilon$ are constant on transverse space ${\mb{R}}^{8} = {\mb{R}}^4 \times {\mb{R}}^4$. In particular they are constant on the first $\mb{R}^4$ in which we embed $S^3$ which will be part of the M5 brane worldvolume. To translate from the BLG spinor to the M5 brane spinor, we first wish to translate a spinor from Cartesian coordinates on ${\mb{R}}^4 = {\mb{R}}_+ \times S^3$, to Spherical coordinates. This is done by means of a transition matrix $g$ as 
\bea
\psi(x) = \psi(E) = g\psi(e)
\eea
where $E$ denotes the vielbein associates with Cartesian coordinates, and $e$ the vielbein associated with Spherical coordinates, and we may refer to $\psi(E)$ as the Cartesian (BLG) spinor and $\psi(e)$ as the Spherical spinor. Details and explicit formulas regarding this map are collected in the Appendix \ref{B} and the general theory of vielbeins is summarized in Appendix \ref{A}. The transition matrix $g$, which is built out of gamma matrices of ${\mb{R}}^4$, commutes with $\t \Gamma$. The chiralities are therefore not affected by $g$,
\bea
\t \Gamma \psi(E) &=& g \t \Gamma \psi(e)
\eea
We subsequently define an M5 brane spinor $\chi$ by
\bea
\psi(e) &=& U \chi(e)
\eea
where
\bea
U &=& \frac{1}{\sqrt{2}} (1-\Sigma)\cr
\Sigma &=& \frac{1}{6}\omega^{\alpha\beta\gamma}\Gamma_{\alpha\beta\gamma}
\eea
We have the following useful duality relations
\bea
\Gamma_{\alpha\beta\gamma} &=& \omega_{\alpha\beta\gamma} \Sigma\cr
\Gamma_{\alpha\beta} &=& \omega_{\alpha\beta\gamma} \Sigma \Gamma^{\gamma}\cr
\Gamma_{\alpha} &=& -\frac{1}{2}\omega_{\alpha\beta\gamma} \Sigma \Gamma^{\beta\gamma}
\eea
We may note that even though $\Gamma_{\alpha}$ are covariantly constant on both $S^3$ and ${\mb{R}}^4$, the same is not true for $\omega_{\alpha\beta\gamma}$, which is covariantly constant only on $S^3$. It means that $D_{\alpha}^{{\mb{R}}^4} U \neq 0$, but it is true that $D_{\alpha}^{(S^3)} U = 0$. Given the above definitions, the M5 brane spinor will be subject to the chirality condition
\bea
\t \Gamma \Sigma \chi &=& \chi
\eea

\subsection{Deconstructing M5 brane Lagrangian}
The mass-deformed sextic potential can be thought of as the sum of three contributions, the undeformed sextic potential, the flux term and the mass term. We list these terms below, expanded around the vacuum to quadratic order (more details on this computation are found in \cite{Gustavsson:2009qd}, though there is some crucial sign error in this work),
\bea
\L_{pot} &=& - \frac{1}{2} (D_{\alpha}Y^{\alpha})^2 - \frac{1}{R^2} Y^{\alpha} Y_{\alpha}\cr
&& + \(\frac{3}{2} - 3^2\) \(\frac{Y}{R}\)^2\cr
&& - 6 \frac{Y}{R} D_{\alpha} Y^{\alpha}\cr
&& - \frac{1}{2} g^{\alpha\beta} \partial_{\alpha} Y \partial_{\beta} Y - \frac{1}{2R^2} Y^{\alpha} Y_{\alpha} 
\eea
\bea
\L_{flux} &=& 4 \frac{Y}{R} D_{\alpha}Y^{\alpha} + 6 \(\frac{Y}{R}\)^2 + \frac{2}{R^2} Y^{\alpha}Y_{\alpha}
\eea
\bea
\L_{m} &=& -\frac{1}{2R^2} Y^{\alpha} Y_{\alpha} - \frac{1}{2} \(\frac{Y}{R}\)^2
\eea
Summing these contributions we get the following contribution from the mass-deformed sextic potential 
\bea
\L_{mpot} &=& -\frac{1}{2} (D_{\alpha}Y^{\alpha})^2 - 2 \frac{Y}{R} D_{\alpha}Y^{\alpha} - 2 \(\frac{Y}{R}\)^2 - \frac{1}{2} g^{\alpha\beta} \partial_{\alpha} Y \partial_{\beta} Y
\eea
In particular we see that the mass term for $Y^{\alpha}$ vanishes by a fortunate cancelation. We dualize
\bea
Y^{\alpha} &=& \frac{1}{2} \omega^{\alpha\beta\gamma}B_{\beta\gamma}
\eea
and define
\bea
D_{\alpha} Y^{\alpha} &=& \frac{1}{6} \omega^{\alpha\beta\gamma} H_{\alpha\beta\gamma}\cr
H_{\alpha\beta\gamma} &=& 3 D_{[\alpha} B_{\beta\gamma]}
\eea
Then we get
\bea
\L_{mpot} &=& - \frac{1}{12} H_{\alpha\beta\gamma}H^{\alpha\beta\gamma} - \frac{1}{3R} Y \omega^{\alpha\beta\gamma}H_{\alpha\beta\gamma}  - 2 \(\frac{Y}{R}\)^2 - \frac{1}{2} g^{\alpha\beta} \partial_{\alpha} Y \partial_{\beta} Y
\eea

Let us proceed to the term
\bea
\frac{i}{4}\bar{\psi} \Gamma_{ij} \{T^i,T^j,\psi\}
\eea
We get
\bea
\bar{\psi}\Gamma_{ij}\omega^{\alpha\beta\gamma}\partial_{\alpha}T^i \partial_{\beta}T^j \partial_{\gamma}\psi
\eea
After making the unitary transformation to M5 brane spinor and noting that $\psi = U \chi$ and $\bar{\psi} = \bar{\chi} U$, we get 
\bea
\frac{i}{2}\bar{\chi} \Gamma^{\alpha} D_{\alpha}^{(S^3)} \chi + \frac{3i}{4R} \bar{\chi} \Sigma \Gamma_R \chi
\eea
We also have a fermionic mass term in mass deformed BLG theory,
\bea
\frac{im}{2} \bar{\psi} \Sigma \Gamma_R \psi
\eea
which becomes 
\bea
-\frac{i}{2R} \bar{\chi} \Sigma \Gamma_R \chi
\eea
Thus the sum of these two contributions becomes
\bea
\frac{i}{2}\bar{\chi} \Gamma^{\alpha} D_{\alpha}^{(S^3)} \chi + \frac{i}{4R} \bar{\chi} \Sigma \Gamma_R \chi
\eea
We also have the kinetic term in BLG theory, which to quadratic order is
\bea
\frac{i}{2}\bar{\psi}\Gamma^{\mu}\partial_{\mu}\psi
\eea
Noting that $g^{-1} \Gamma^{\mu} g = \Gamma^{\mu}$ (the $g$ commutes with $\Gamma_{\mu}$) and $U \Gamma^{\mu} U = \Gamma^{\mu}$, this term becomes
\bea
\frac{i}{2}\bar{\chi}\Gamma^{\mu}\partial_{\mu}\chi
\eea
To summarize, the fermionic part of the deconstructed M5 brane Lagrangian read
\bea
\frac{i}{2}\(\bar{\chi}\Gamma^{\mu}\partial_{\mu}\chi + \bar{\chi} \Gamma^{\alpha} D_{\alpha}^{(S^3)} \chi\) + \frac{i}{4R} \bar{\chi} \Sigma \Gamma_R \chi 
\eea
and the fermionic equation of motion becomes
\bea
\Gamma^{\mu}\partial_{\mu}\chi + \Gamma^{\alpha} D_{\alpha}^{(S^3)} \chi + \frac{1}{2R} \Sigma \Gamma_R \chi  &=& 0
\eea

We include kinetic terms, the CS term, additional mass terms for $Y^{\hat{i}}$, and add the fermions, and we define
\bea
H_{\mu \alpha\beta} &:=& \partial_{\mu}B_{\alpha\beta} + D_{\alpha} B_{\beta\mu} - D_{\beta}B_{\alpha\mu}\cr
B_{\alpha\mu} &:=& -B_{\mu\alpha}
\eea
and we obtain the full M5 brane Lagrangian as\footnote{Gauge invariance in the first line of this Lagrangian is not manifest. However, by adding certain total derivative terms, the first line can be brought into the form \cite{Pasti:2011zz}
\bea
 - \frac{1}{2} H_{\mu\alpha\beta} {H^{-}}^{\mu\alpha\beta} - \frac{1}{6} H_{\alpha\beta\gamma} {H^{-}}^{\alpha\beta\gamma}
\eea
where 
\bea
{H^{-}}^{\mu\alpha\beta} &=& \frac{1}{2} \(H^{\mu\alpha\beta} - \frac{1}{2}\epsilon^{\mu\nu\lambda}\omega^{\alpha\beta\gamma} H_{\nu\lambda\gamma}\)\cr
{H^{-}}^{\alpha\beta\gamma} &=& \frac{1}{2} \(H^{\alpha\beta\gamma} - \frac{1}{6}\epsilon^{\mu\nu\lambda}\omega^{\alpha\beta\gamma} H_{\mu\nu\lambda}\)
\eea
and 
\bea
H_{\mu\nu\lambda} &=& \partial_{\mu}B_{\nu\lambda} + \partial_{\lambda} B_{\mu\nu} + \partial_{\nu} B_{\lambda\mu}\cr
H_{\mu\nu\alpha} &=& \partial_{\mu} B_{\nu\alpha} + \partial_{\alpha} B_{\mu\nu} + \partial_{\nu} B_{\alpha\mu}
\eea
It should be noted that we now have introduced gauge field components $B_{\mu\nu}$. A careful analysis reveals that these all cancel up to total derivative terms. So all we have done is really nothing but adding certain total derivative terms to the Lagrangian (\ref{Lagr}). }
\ben
\L &=& -\frac{1}{12} H_{\alpha\beta\gamma}H^{\alpha\beta\gamma} - \frac{1}{4} H_{\mu\alpha\beta} H^{\mu\alpha\beta} + \frac{1}{2} \epsilon^{\mu\nu\lambda} \omega^{\alpha\beta\gamma} \partial_{\beta} B_{\mu\alpha} \partial_{\nu}B_{\lambda\gamma}\cr
&& - \frac{1}{2} g^{\alpha\beta} \partial_{\alpha} Y^A \partial_{\beta} Y^A - \frac{1}{2} \partial_{\mu} Y^A \partial^{\mu} Y^A\cr
&& - \frac{1}{3R} Y \omega^{\alpha\beta\gamma}H_{\alpha\beta\gamma}  - \frac{2}{R^2} Y^2 - \frac{1}{2R^2} Y^{\hat{i}}Y^{\hat{i}}\cr
&& +\frac{i}{2}\(\bar{\chi}\Gamma^{\mu}\partial_{\mu}\chi + \bar{\chi} \Gamma^{\alpha} D_{\alpha}^{(S^3)} \chi\) + \frac{i}{4R} \bar{\chi} \Sigma \Gamma_R \chi \label{Lagr}
\een
but we have not yet fixed the overall normalization, which determines the M5 brane coupling constant. We will determine the coupling constant in section \ref{coupling}.

\subsection{Deconstructing M5 supersymmetry variations}
In Appendix \ref{killing}, eq (\ref{killingeq}), we find that the BLG supersymmetry parameter satisfies the Killing spinor equation
\bea
D_{\alpha}^{(S^3)} \epsilon &=& -\frac{1}{2R}\Gamma_{\alpha}\Gamma_R\epsilon
\eea
We now map this to the corresponding equation for the M5 brane supersymmetry parameter $\omega$ that we shall define by
\bea
\epsilon &=& U\omega
\eea
The chiralities are such that
\bea
\t \Gamma \epsilon &=& \epsilon\cr
\t \Gamma \Sigma \omega &=& - \omega
\eea
for M2 and M5 brane SUSY parameters respectively. These chiralities are related by the unitary transformation matrix $U$. Recalling that $D_{\alpha}^{(S^3)} U = 0$, we get
\bea
D_{\alpha}^{(S^3)} \omega &=& -\frac{1}{2R}\Gamma_{\alpha}\Sigma \Gamma_R \omega
\eea
where we used
\bea
U^{-1}\Gamma_{\alpha}\Gamma_R U &=& \Gamma_{\alpha}\Sigma \Gamma_R
\eea
To close supersymmetry, we will need the result
\bea
D_{\alpha}^{(S^3)} \bar{\omega} &=& \frac{1}{2R} \bar{\omega} \Sigma \Gamma_R \Gamma_{\alpha}
\eea

With the above specified chirality on SUSY parameters, these will be subject to the Fierz identity in $1+5$ dimensions, where $M=(\mu,\alpha)$ \footnote{
For clarity we display only the part of the Fierz identity which is valid when we act on some chiral spinor so that $\t \Gamma \Sigma $ is replaced by $1$.}
\bea
\omega \bar{\rho} - \rho \bar{\omega} = - \frac{1}{8} (\bar{\rho}\Gamma_M \omega)\Gamma^M + \frac{1}{8} (\bar{\rho}\Gamma_M\Gamma_A\omega)\Gamma^M\Gamma^A - \frac{1}{192} (\bar{\rho}\Gamma_{MNP}\Gamma_{AB}\omega)\Gamma^{MNP}\Gamma^{AB}
\eea
which we will use in the following form
\bea
\omega \bar{\rho} - \rho \bar{\omega} &=& - \frac{1}{8} (\bar{\rho}\Gamma_M \omega)\Gamma^M + \frac{1}{8} (\bar{\rho}\Gamma_M\Gamma_A\omega)\Gamma^M\Gamma^A \cr
&& - \frac{1}{16} (\bar{\rho}\Sigma\Gamma_{AB}\omega)\Sigma\Gamma^{AB} - \frac{1}{16} (\bar{\rho}\Sigma \Gamma_{\alpha} \Gamma_{\mu} \Gamma_{AB}\omega) \Sigma \Gamma^{\alpha}\Gamma^{\mu}\Gamma^{AB}
\eea

\subsubsection*{Zeroth order}
As we have already mentioned, by choosing the background as $S^3$ with radius given by Eq (\ref{m}), the supersymmetry variation of the fermion vanishes to zeroth order in the fluctuations. It is important to note that this happens for any choice of supersymmetry parameter $\epsilon$ and does not give us any Weyl projection condition on $\epsilon$ as usually happens for BPS solutions. So the $S^3$ background preserves maximal supersymmetry. 

\subsubsection*{Linear order -- cancelation of gauge field mass term}
As another consequence of $m = -\frac{1}{R}$ we find that the mass term for the gauge field cancels. Having already canceled out the zeroth order contribution, the explicit mass term at linear order in BLG theory reads 
\bea
\delta_m \psi &=& -m\Gamma_{(4)} \Gamma_I \epsilon Y^I
\eea
We expand $Y^I$ and we rotate the spinors by the transition matrix $g$ into polar coordinates, and we get
\bea
\delta_m \psi &=& - m \Sigma \Gamma_R \Gamma_{\alpha} \epsilon Y^{\alpha} - m \Sigma \Gamma_R \Gamma_A \epsilon Y^A
\eea
We get further mass terms from expanding the sextic potential term
\bea
\delta_{pot} \psi &=& - \frac{1}{R} \Sigma \Gamma_R \Gamma_{\alpha} \epsilon Y^{\alpha} - 3 \Sigma \epsilon \frac{Y}{R}
\eea
We see that the mass term of the gauge field cancels and the sum becomes
\bea
\delta \psi &=& \frac{1}{R} \Sigma \Gamma_R \Gamma_A \epsilon Y^A - 3 \Sigma \epsilon \frac{Y}{R}
\eea
We map to M5 quantities and get
\bea
\delta \chi &=& \frac{1}{R} \Sigma \Gamma_R \Gamma_A \omega Y^A - 3 \Sigma \omega \frac{Y}{R}
\eea
In order to avoid the technical problem of having to think on a sign when we commute $\Gamma_R$ with $\Gamma_A$, we will write this same variation in the form
\bea
\delta \chi &=& \frac{1}{R} \Sigma \Gamma_R \Gamma_{\hat{i}} \omega Y^{\hat{i}} - 2 \Sigma \omega \frac{Y}{R}
\eea

We thus arrive at the following M5 brane SUSY variations
\bea
\delta Y_A &=& i \bar{\omega} \Gamma_A \chi\cr
\delta B_{\mu\alpha} &=& i \bar{\omega} \Gamma_{\mu}\Gamma_{\alpha} \chi\cr
\delta B_{\alpha\beta} &=& i \bar{\omega} \Gamma_{\alpha\beta} \chi\cr
\delta \chi &=& \frac{1}{2} \Gamma^{\mu}\Gamma^{\alpha\beta} \omega H_{\mu\alpha\beta} + \frac{1}{6} \Gamma^{\alpha\beta\gamma}\omega H_{\alpha\beta\gamma}\cr
&&+\Gamma^{\mu}\Gamma^A \omega \partial_{\mu} Y_A + \Gamma^{\alpha}\Gamma^A \omega \partial_{\alpha} Y_A\cr
&&-\frac{2}{R} \Sigma \omega Y + \frac{1}{R} \Sigma \Gamma_R \Gamma_{\hat{i}} \omega Y^{\hat{i}}
\eea

\subsection{Closure on fermions}
Defining
\bea
v^M &=& \bar{\rho}\Gamma^M \omega
\eea
we have to zeroth order in $\frac{1}{R}$ the closure relations
\bea
[\delta_{\rho},\delta_{\omega}]\chi &=& 2i v^{\nu} \partial_{\nu} \chi + 2i v^{\delta} D_{\delta} \chi\cr
&& - i v^{\nu} \Gamma_{\nu} \slashed{D} \chi\cr
&& - \frac{i}{2} v^{\delta} \Gamma_{\delta} \slashed{D} \chi\cr
&& + \frac{i}{2} (\bar{\rho} \Gamma_{\alpha} \Gamma_A \omega) \Gamma^{\alpha} \Gamma^A \slashed{D} \chi\cr
&& + \frac{i}{4} (\bar{\rho} \Sigma \Gamma_{AB} \omega) \Gamma^{AB} \slashed{D}\chi
\eea
The $\frac{1}{R}$ corrections contribute
\bea
[\delta_{\rho},\delta_{\omega}] \chi &=& \Big(-\frac{i}{2R}(\bar{\rho}\Gamma^{\mu}\omega)\Gamma_{\mu} - \frac{i}{4R}(\bar{\rho}\Gamma^{\alpha}\omega)\Gamma_{\alpha}\cr
&& + \frac{5i}{4R}(\bar{\rho}\Gamma_R \Gamma^{\alpha}\omega)\Gamma_R\Gamma_{\alpha}\cr
&& + \frac{i}{4R}(\bar{\rho}\Gamma_{\hat{i}}\Gamma^{\alpha}\omega)\Gamma^{\hat{i}}\Gamma_{\alpha}\Big)\Sigma\Gamma_R\chi
\eea
Adding up these contributions, we find
\bea
[\delta_{\rho},\delta_{\omega}] \chi &=& 2i v^{\mu} \partial_{\mu} \chi + 2i \L_v \chi\cr
&& - i v^{\mu} \Gamma_{\mu} \(\slashed{D} \chi + \frac{1}{2R} \Sigma \Gamma_R\chi\)\cr
&& - \frac{i}{2} v^{\alpha} \Gamma_{\alpha} \(\slashed{D} \chi + \frac{1}{2R} \Sigma \Gamma_R \chi\)\cr
&& + \frac{i}{2} (\bar{\rho} \Gamma_{\alpha}\Gamma_A \omega) \Gamma^{\alpha} \Gamma^A \(\slashed{D} \chi + \frac{1}{2R} \Sigma \Gamma_R \chi\)\cr
&& - \frac{i}{2} (\bar{\rho} \Sigma \Gamma_R \Gamma_{\hat{i}} \omega) \Sigma \Gamma^R \Gamma^{\hat{i}} \(\slashed{D}\chi + \frac{1}{2R} \Sigma \Gamma_R \chi\)\cr
&& - \frac{i}{4} (\bar{\rho} \Sigma \Gamma_{\hat{i}\hat{j}} \omega) \Sigma \Gamma^{\hat{i}\hat{j}} \(\slashed{D}\chi + \frac{1}{2R} \Sigma \Gamma_R \chi\)\cr
&& + \frac{i}{2R} (\bar{\rho} \Sigma \Gamma_{\hat{i}\hat{j}} \omega) \Gamma^{\hat{i}\hat{j}} \Gamma_R \chi
\eea
where
\bea
\L_{v} \chi &=& v^{\alpha} D_{\alpha} \chi - \frac{1}{2R} (\bar{\rho} \Gamma_R \Gamma_{\alpha} \omega) \Gamma^{\alpha} \Sigma \chi
\eea
indeed is the Lie derivative. To see this we compute
\bea
D_{\alpha} v_{\beta} &=& - \frac{1}{R}\omega_{\alpha\beta\gamma} \bar{\rho} \Gamma_R \Gamma^{\gamma} \omega
\eea
Here we used the Killing spinor equation for $\rho$ and $\omega$ in the definition $v_{\beta} = \bar{\rho}\Gamma_{\beta}\omega$. We present the expression of the Lie derivative in eq (\ref{Lieder}) in Appendix \ref{Lie}.

The last term is an $SO(4)$ R symmetry rotation, and we will return to this term in the next subsection.

\subsection{Closure on bosons}
For the scalars we get the closure relation
\bea
[\delta_{\rho},\delta_{\omega}]Y_{\hat{i}} &=& 2i v^{\mu} \partial_{\mu} Y_{\hat{i}} + 2i \L_v Y_{\hat{i}} + \frac{2i}{R}\bar{\omega}\Gamma_{\hat{i}\hat{j}} \Sigma \Gamma_R \rho Y_{\hat{j}}\cr
[\delta_{\rho},\delta_{\omega}]Y &=& 2i v^{\mu} \partial_{\mu} Y + 2i \L_v Y
\eea
We can bring the last term into the form
\bea
-\frac{i}{R}\epsilon_{\hat{i}\hat{j}\hat{k}\hat{l}}(\bar{\rho}\Gamma_{\hat{k}\hat{l}}\Sigma\omega) Y_{\hat{j}}
\eea
which should be compared to the corresponding term for the fermion
\bea
\frac{i}{2R} (\bar{\rho} \Sigma \Gamma_{\hat{i}\hat{j}} \omega) \Gamma^{\hat{i}\hat{j}} \Gamma_R \chi &=& -\frac{i}{4R}\epsilon_{\hat{i}\hat{j}\hat{k}\hat{l}} (\bar{\rho} \Sigma \Gamma_{\hat{i}\hat{j}} \omega) \Gamma^{\hat{k}\hat{l}} \chi 
\eea
We can identify both these terms as an $SO(4)$ R-symmetry rotation $\delta_R$ which acts on the scalars and the fermions according to
\bea
\delta_R Y_{\hat{i}} &=& -\frac{i}{2R} \epsilon_{\hat{i'}\hat{j'}\hat{k}\hat{l}} (\bar{\rho} \Sigma \Gamma_{\hat{i'}\hat{j'}} \omega) (M_{\hat{k}\hat{l}})_{\hat{i}\hat{j}} Y_{\hat{j}}\cr
\delta_R \chi &=& -\frac{i}{2R} \epsilon_{\hat{i'}\hat{j'}\hat{k}\hat{l}} (\bar{\rho} \Sigma \Gamma_{\hat{i'}\hat{j'}} \omega) M_{\hat{k}\hat{l}} \chi
\eea
where $M_{\hat{i}\hat{j}}$ is a generator of $SO(4)$ in the vector and spinor representation respectively (see section Appendix A.1 for our conventions).

For the gauge potential we get
\bea
[\delta_{\rho},\delta_{\omega}]B_{\mu\alpha} &=& 2i v^{\gamma} H_{\mu\alpha\gamma} - i v^{\lambda} \epsilon_{\mu\nu\lambda}\omega_{\alpha}{}^{\beta\gamma} H_{\nu\beta\gamma}
\eea
\bea
[\delta_{\rho},\delta_{\omega}]B_{\alpha\beta} &=& 2i v^{\nu} H_{\nu\alpha\beta} + 2i v^{\gamma} H_{\alpha\beta\gamma} + \Lambda_{\alpha\beta}\cr
\Lambda_{\alpha\beta} &=& \frac{4i}{R} \omega_{\alpha\beta\gamma} v^{\gamma} Y
\eea
The gauge invariant field strength is given by
\bea
\t H_{\alpha\beta\gamma} &=& 3 D_{[\alpha} B_{\beta\gamma]} + C_{\alpha\beta\gamma}
\eea
where $C_{\alpha\beta\gamma}$ is the background three-form gauge potential in eleven-dimensional supergravity. From our deconstruction we obtain
\ben
C_{\alpha\beta\gamma} &=& -\frac{2}{R} \omega_{\alpha\beta\gamma} Y\label{gaug}
\een
We note that in the action as well as in the supersymmetry variation of the fermion, the combination $\t H_{\alpha\beta\gamma} = H_{\alpha\beta\gamma} - \frac{2}{R}\omega_{\alpha\beta\gamma} Y$ always appears, in the action this combination appears squared. The gauge symmetry should now acts as
\bea
\delta B_{\alpha\beta} &=& \Lambda_{\alpha\beta} + 2 \partial_{[\alpha}\Lambda_{\beta]}\cr
\delta C_{\alpha\beta\gamma} &=& - 3\partial_{[\alpha}\Lambda_{\beta\gamma]}
\eea
which leaves the combination $\t H_{\alpha\beta\gamma}$ invariant. We notice that since $Y$ is gauge invariant, Eq (\ref{gaug}) is gauge non-covariant. Somewhere in our deconstruction it thus appears as if we have made some sort of gauge fixing, which we have not been aware of. Presumably this can be attributed to a the fact that we have been ignorant about a residual shift symmetry of the fermion in BLG theory. That shift symmetry presumably transmutes in the M5 brane into a gauge variation of $C_{\alpha\beta\gamma}$. Such a gauge transformation may depend on $\sigma^{\alpha}$ since what matters is that the shift of the fermion in BLG theory does not depend on space-time of BLG theory, that is, on $x^{\mu}$. Hence we may still have an arbitrary $\sigma^{\alpha}$ dependence on the `constant' shift of the BLG fermion.

We can write these closure relations as
\bea
[\delta_{\rho},\delta_{\omega}]B_{\alpha\beta} &=& 2i v^{\nu} H_{\nu\alpha\beta} + 2i \L_v B_{\alpha\beta} \cr
&& + \partial_{\alpha}\Lambda_{\beta} - \partial_{\beta} \Lambda_{\alpha} + \Lambda_{\alpha\beta}\cr
[\delta_{\rho},\delta_{\omega}]B_{\mu\alpha} &=& 2i \L_v B_{\mu\alpha} + \partial_{\mu}\Lambda_{\alpha}-D_{\alpha}\Lambda_{\mu}\cr
\Lambda_{\mu} &=& v^{\beta} B_{\beta\mu}\cr
\Lambda_{\alpha} &=& v^{\beta} B_{\beta\alpha}\cr
\Lambda_{\alpha\beta} &=& \frac{4i}{R} \omega_{\alpha\beta\gamma} v^{\gamma} Y
\eea
where in this case the Lie derivative on $S^3$ is given by
\bea
\L_v B_{\alpha\beta} &=& v^{\gamma} D_{\gamma} B_{\alpha\beta} + D_{\alpha} v^{\gamma} B_{\gamma\beta} + D_{\beta} v^{\gamma} B_{\alpha\gamma}
\eea

We have also performed a check of the supersymmetry of the action. This computation is summarized in Appendix \ref{M5close}.

\section{Dimensional reduction}
We first develop general formalism for dimensional reduction on a circle. We consider a generic circle-bundle $M_3$ over a two-manifold $M_2$. (We will later take as $M_3 = S^3$ and $M_2 = S^2$, but for now our discussion will be general). The most general metric on $M_3$, which is translationally invariant along the circle, can be written as
\bea
ds^2 &=& G_{mn} d\sigma^m d\sigma^n + g_{\psi\psi} (d\psi + V_m d\sigma^m)^2
\eea
where $\sigma^m $ parameterize $M_2$, and $\psi \in [0,2\pi]$ is the coordinate of the circle. Here $V_m$ is a connection one-form that is associated with the twisting of the circle-bundle. If we gather the coordinates as $\sigma^{\alpha} = (\sigma^m, \psi)$, the metric tensor thus has components
\ben
g_{\alpha\beta} &=& \(\begin{array}{cc}
G_{mn} + g_{\psi\psi} V_m V_n & g_{\psi\psi} V_m\\
g_{\psi\psi} V_n & g_{\psi\psi}
\end{array}\)\label{KKmetric}
\een
and this can be inverted as
\bea
g^{\alpha\beta} &=& \(\begin{array}{cc}
G^{mn} & -V^m\\
-V^n & \frac{1}{g_{\psi\psi}} + V^2
\end{array}\)
\eea
Here
\bea
V^2 &:=& G^{mn} V_m V_n
\eea
We choose the vielbein as
\bea
e_{\alpha}{}^i &=& \(\begin{array}{cc}
E_m^I & \sqrt{g_{\psi\psi}} V_m\\
0 & \sqrt{g_{\psi\psi}}
\end{array}\)
\eea
\bea
e_i{}^{\alpha} &=& \(\begin{array}{cc}
E_I^m & -V_I\\
0 & \frac{1}{\sqrt{g_{\psi\psi}}}
\end{array}\)
\eea
where
\bea
V_I &=& E_I^m V_m
\eea
and we let $i = (I,3)$ where $I=1,2$. Given a basis for flat space gamma matrices $\Gamma_i = (\Gamma_I,\Gamma_3)$ which obey $\{\Gamma_i,\Gamma_j\} = 2\delta_{ij}$, we define 
\bea
\Gamma_{\alpha} &=& e_{\alpha}^i \Gamma_i\cr
\t \Gamma_m &=& E_m^I \Gamma_I
\eea
and we find that 
\bea
\Gamma_m &=& \t \Gamma_m + V_m \Gamma_{\psi}
\eea
and
\ben
\Gamma^m &=& \t \Gamma^m\cr 
\Gamma^{\psi} &=& \frac{1}{g_{\psi\psi}} \Gamma_{\psi} - V^m \t \Gamma_m \label{gammapsi}
\een
We have $\{\t \Gamma_m,\Gamma_{\psi}\} = 0$ whereas $\{\Gamma_m,\Gamma_{\psi}\} = 2 g_{\psi\psi} V_m$.

If we define
\bea
G &:=& \det G_{mn}\cr
g &:=& \det g_{\alpha\beta}
\eea
then we have the following relation between the totally antisymmetric tensors in two and three dimensions,
\bea
\omega_{mn\psi} &=& \sqrt{\frac{g}{G}} \omega_{mn}\cr
\omega^{mn\psi} &=& \sqrt{\frac{G}{g}} \omega^{mn}
\eea
Associated with these, we define
\bea
\Sigma &=& \frac{1}{6}\omega^{\alpha\beta\gamma}\Gamma_{\alpha\beta\gamma}\cr
\t \sigma &=& \frac{1}{2}\omega^{mn}\t \Gamma_{mn}
\eea
We then get
\bea
\frac{1}{6} \omega^{\alpha\beta\gamma} \Gamma_{\alpha\beta\gamma} &=& \sqrt{\frac{G}{g}} \t \sigma \Gamma_{\psi}\cr
\frac{1}{6} \omega_{\alpha\beta\gamma} \Gamma^{\alpha\beta\gamma} &=& \sqrt{\frac{g}{G}} \frac{1}{g_{\psi\psi}} \t \sigma \Gamma_{\psi}
\eea
Of course these results must agree as we just rised and lowered indices by $g_{\alpha\beta}$, and therefore we must have the identity
\bea
g &=& g_{\psi\psi} G
\eea
Indeed this can be verified directly. Not so easily by looking at the Kaluza-Klein metric (\ref{KKmetric}), but we can look at the vielbeins and see that
\bea
\det e_{\alpha}^i &=& \sqrt{g_{\psi\psi}} \det E_m^I
\eea
We conclude that
\bea
\Sigma &=& \frac{1}{\sqrt{g_{\psi\psi}}} \t \sigma \Gamma_{\psi}
\eea
Let us henceforth abbreviate
\bea
g_{\psi\psi} &=& R^2
\eea

We will also need the dimensional reduction of the covariant derivative. To this end we consider the reduction of the spin connection. The spin connections in three and two dimensions satisfy
\bea
de^i + \omega^{ij} \wedge e^j &=& 0\cr
dE^I + \Omega^{IJ} \wedge E^J &=& 0
\eea
on $M_3$ and $M_2$ respectively. We split $i = (I,3)$ and write the first equation as
\bea
de^I + \omega^{IJ} \wedge e^J + \omega^{I3} \wedge e^3 &=& 0
\eea
We define 
\bea
W_{mn} &=& \partial_m V_n - \partial_n V_m
\eea
We then find that a solution to these equation is given by 
\bea
\omega^{IJ}_m &=& \Omega^{IJ}_m - \frac{R}{2} W^{IJ} \(RV_m + \partial_m \psi\)\cr
\omega^{I3}_m &=& \frac{R}{2} W_{mn} E^{In}\cr
\omega^{IJ}_{\psi} &=& - \frac{R^2}{2} W^{IJ}
\eea
The relation between $M_3$ and $M_2$ covariant derivatives then becomes
\bea
D_m^{S^3} &=& D_m^{S^2} - \frac{R^2}{8} V_m W^{pq} \t \Gamma_{pq} + \frac{R}{4} W_{mn} \t \Gamma^n \Gamma_3 \cr
D_{\psi}^{S^3} &=& \partial_{\psi} - \frac{R^2}{8} W_{mn} \t \Gamma^{mn} 
\eea

\subsection{Dimensional reduction of supersymmetry}
Let us first dimensionally reduce the Killing spinor equation for the supersymmetry parameter. On the Hopf bundle $S^3 \rightarrow S^2$ we have
\bea
W_{mn} &=& -\frac{2}{R^2} \omega_{mn}
\eea
and we find 
\bea
D_m^{S^3} &=& D_m^{S^2} + \frac{1}{2} V_m \t \sigma - \frac{1}{2R^2} \t \Gamma_m \t \sigma \Gamma_{\psi}\cr
D_{\psi}^{S^3} &=& \partial_{\psi} + \frac{1}{2} \t \sigma
\eea
On the M5 the spinor and the supersymmetry parameter have opposite six-dimensional chiralities 
\bea
\Gamma \omega &=& -\omega\cr
\Gamma \chi &=& \chi
\eea
For five-dimensional super Yang-Mills we shall have chiral spinor and a chiral supersymmetry parameter, but with the same chiralities. There are no chiral spinors in five dimensions. But if we use ten-dimensional spinors, then we shall impose ten-dimensional chirality. Since we are coming from eleven-dimensional spinors and we wish to get rid of the $\psi$-direction to descend to ten dimensions, the natural choice for the chirality matrix from a ten-dimensional viewpoint is $\frac{1}{R}\Gamma_{\psi}$. We divide by $R$ because $\Gamma_{\psi}^2 = g_{\psi\psi} = R^2$. We thus wish to work with spinor $\psi$ and supersymmetry parameter $\epsilon$ subject to chirality conditions
\bea
\frac{1}{R} \Gamma_{\psi} \epsilon &=& \epsilon\cr
\frac{1}{R} \Gamma_{\psi} \psi &=& \psi
\eea
To this end, we define 
\bea
\omega &=& u^{\dag} \epsilon\cr
\chi &=& c u \psi
\eea
where $c$ is a normalization and
\bea
u &=& \frac{1}{\sqrt{2}} (1+\gamma)\cr
\gamma &=& \t \Gamma \t \sigma
\eea
where
\bea
\frac{1}{R}\Gamma_{\psi} &=& \Gamma_{012} \t \sigma \Gamma_{\hat{1}\hat{2}\hat{3}\hat{4}}\cr
\t \sigma &=& \frac{1}{2} \omega^{mn}\t \Gamma_{mn}
\eea
Contracting by $\Gamma_{\psi}$ on both sides, we get
\bea
1 &=& \t \Gamma \Sigma \hat{\Gamma}
\eea
where we used 
\bea
\Sigma &=& \frac{1}{R} \t \sigma \Gamma_{\psi} 
\eea

From $\Gamma \omega = -\omega$ we then get the condition that
\bea
\(\gamma - 1\)\(\frac{\Gamma_{\psi}}{R} - 1\) \epsilon &=& 0
\eea
As the operator $\gamma - 1 \neq 0$ we get the desired chirality condition $\frac{\Gamma_{\psi}}{R} \epsilon = \epsilon$. Similar type of computation applies to $\psi$. 

To dimensionally reduce the M5 supersymmetries, we also need the result 
\bea
\bar{\omega} &=& \bar{\epsilon} u
\eea
Let us collectively denote five-dimensional indices by
\bea
M &=& (\mu,m)
\eea
We have commutation relations
\bea
[\gamma, \Gamma_M] &=& 0\cr
\{\gamma, \Gamma_A\} &=& 0\cr
\{\gamma, \Gamma_{\psi}\} &=& 0
\eea

Let us now dimensionally reduce the Killing spinor equation which corresponds to a constant Dirac spinor in ${\mb{R}}^4$. In spherical coordinates this spinor, which we may denote as $\E$, satisfies the Killing spinor equation 
\bea
D^{S^3}_{\alpha} \E &=& - \frac{1}{2R} \Gamma_{\alpha}\Gamma_R \E
\eea
on $S^3$. Let us first consider the $\psi$ component of this equation, 
\bea
D_{\psi}^{S^3} \E &=& -\frac{1}{2R} \Gamma_{\psi}\Gamma_R \E
\eea
Dimensional reduction amounts to letting
\bea
\partial_{\psi} \E &=& 0
\eea
and then the $\psi$ component of the Killing spinor equation reduces to a Weyl projection on ${\mb{R}}^4$,
\ben
\t \sigma \E &=& - \frac{\Gamma_{\psi}}{R} \Gamma_R \E\label{chirality}
\een
We next turn to the $S^2$ components of the Killing spinor equation,
\bea
D_m^{S^3} \E &=& -\frac{1}{2R} \Gamma_m \Gamma_R \E
\eea
which, by expanding out the covariant derivative, reads
\bea
D_m^{S^2} \E + \frac{1}{2} V_m \t \sigma \E - \frac{1}{2R^2} \t \Gamma_m \t \sigma \Gamma_{\psi} \E &=& -\frac{1}{2R} \Gamma_m \Gamma_R \E
\eea
Making the replacement $\Gamma_m = \t \Gamma_m + V_m \Gamma_{\psi}$ in the right-hand side, and using (\ref{chirality}) we see that the terms involving the graviphoton cancel. Then, by again using (\ref{chirality}), we find that two remaining terms add up, and therefore we descend to the Killing spinor equation on $S^2$ with radius $\frac{R}{2}$,
\bea
D_m^{S^2} \E &=& -\frac{1}{R} \t \Gamma_m \Gamma_R \E
\eea
The possible choices for what the right-hand side could be is also very restricted by curvature constraint when commuting two covariant derivatives. 

We may embed the four dimensional spinor $\E$ into the eleven-dimensional BLG supersymmetry parameter $\epsilon$, and then map this to M5 brane spinor $\epsilon = U \omega$ and subsequently to D4 brane spinor $\omega = u^{\dag} \epsilon$. Following this chain, we find for the D4 brane spinor the Weyl conditions
\bea
\epsilon &=& \gamma \Gamma_R \epsilon\cr
\epsilon &=& \frac{\Gamma_{\psi}}{R} \epsilon
\eea
Since $\gamma^2 = -1$, the first Weyl projection can be expressed in the form
\ben
\(\Gamma_R + \gamma\) \epsilon &=& 0
\een
This spinor satisfies the Killing spinor equation
\bea
D_{m}^{S^2} \epsilon &=& \frac{1}{R}\t \Gamma_m \t \sigma \Gamma_R \epsilon
\eea

We define the dimensionally reduced five-dimensional gauge field as
\bea
A_M &=& \int_0^{\frac{2\pi}{k}} d\psi B_{M\psi}
\eea
We reduce the radial component of the scalar field as
\bea
\Phi &=& 2\pi \frac{R}{k} Y
\eea
and likewise for the other components we define
\bea
\phi^{\hat{i}} &=& 2\pi \frac{R}{k} Y^{\hat{i}}
\eea
For the fermions we define
\bea
\psi &=& 2\pi \frac{R}{k} u^{\dag} \chi 
\eea

We are now ready to dimensionally reduce the M5 supersymmetries. For the scalar variations we simply get
\bea
\delta \phi_A &=& i \bar{\epsilon} \Gamma_A \psi
\eea
For the gauge potential variation we get
\bea
\delta A_M &=& \frac{i}{R} \bar{\epsilon} \Gamma_M \Gamma_{\psi} \psi
\eea
By a subsequent use of the chirality condition, this is reduced to
\bea
\delta A_M &=& i \bar{\epsilon} \Gamma_M \psi
\eea

We finally turn to the variation of the fermions. First let us consider the contribution from the gauge field only, that is the terms
\bea
\delta \chi &=& \frac{1}{6}\Gamma^{\alpha\beta\gamma}\omega H_{\alpha\beta\gamma} + \frac{1}{2} \Gamma^{\mu}\Gamma^{\alpha\beta} \omega H_{\mu\alpha\beta}
\eea
We define
\ben
H_{mn \psi} &=& \frac{k}{2\pi} F_{mn}\cr
H_{\mu m \psi} &=& \frac{k}{2\pi} F_{\mu m}\cr
H_{\mu m n} &=& \frac{k}{2\pi} \(\frac{\mu}{2} \epsilon_{\mu\nu\lambda} \omega_{mn} F^{\nu\lambda} + \lambda F_{\mu m} V_n\)\label{mu}
\een
along with 
\bea
\psi &=& \frac{2\pi R}{k} u^{\dag} \chi
\eea
We have made an ansatz with free parameters $\mu$ and $\lambda$ that we will now fix. We get
\bea
\delta \psi &=& \frac{1}{2}\t \Gamma^{mn} \epsilon F_{mn} + \Gamma^{\mu} \t \Gamma^m \epsilon F_{\mu m} - \frac{R\mu}{2} \Gamma^{\mu\nu} \epsilon F_{\mu\nu}\cr
&& + \(R - \frac{R\lambda}{2}\) \Gamma^{\mu} \t \Gamma^{mn} \gamma \epsilon F_{\mu m} V_n
\eea
By taking 
\bea
\mu &=& -\frac{1}{R}\cr
\lambda &=& 2
\eea
and if we define $\Gamma^M = (\Gamma^{\mu},\t \Gamma^m)$ we descend to
\bea
\delta \psi &=& \frac{1}{2} \Gamma^{MN} \epsilon F_{MN}
\eea

Let us move on to the other terms
\bea
\delta \chi &=& \Gamma^{\mu} \Gamma^A \omega \partial_{\mu} Y_A + \Gamma^{\alpha} \Gamma^A \omega \partial_{\alpha} Y_A
\eea
becomes, by noting that we put $\partial_{\psi} Y_A = 0$,
\bea
\delta \psi &=& \Gamma^M \Gamma^A \epsilon \partial_M Y_A
\eea
Finally we have the correction terms
\bea
\delta \chi &=& -\frac{2}{R} \Sigma \omega Y + \frac{1}{R} \Sigma \Gamma_R \Gamma_{\hat{i}} \omega Y^{\hat{i}}
\eea
which become
\bea
\delta \psi &=& -\frac{2}{R} \Sigma \epsilon \Phi + \frac{1}{R} \Sigma \Gamma_R \Gamma_{\hat{i}} \epsilon \phi^{\hat{i}}
\eea
which we can also write as
\bea
\delta \psi &=& -\frac{2}{R^2} \t \sigma \Gamma_{\psi}\epsilon \Phi + \frac{1}{R^2} \t \sigma \Gamma_R \Gamma_{\hat{i}} \Gamma_{\psi} \epsilon \phi^{\hat{i}}
\eea
and which becomes by means of chirality condition
\bea
\delta \psi &=& -\frac{2}{R} \t \sigma \epsilon \Phi + \frac{1}{R} \t \sigma \Gamma_R \Gamma_{\hat{i}} \epsilon \phi^{\hat{i}}
\eea

\subsection{Dimensional reduction of M5 Action}
\subsubsection{Dimensional reduction of bosonic terms}
Let us start with dimensionally reducing the Maxwell part of the M5 brane action,
\bea
S_{e.m.} &=& \frac{1}{2\pi}\int d^6 x \sqrt{g} \(-\frac{1}{12} H_{\alpha\beta\gamma} H^{\alpha\beta\gamma} - \frac{1}{4} H_{\mu\alpha\beta} H^{\mu\alpha\beta} + \frac{1}{2} \epsilon^{\mu\nu\lambda} \omega^{\alpha\beta\gamma} \partial_{\beta} B_{\mu\alpha} \partial_{\nu} B_{\lambda\gamma}\)
\eea
where the overall factor of $\frac{1}{2\pi}$ will be derived in section \ref{coupling}. We split $\alpha = (m,\psi)$, and put $\partial_{\psi} = 0$ and put $f_{MN} = \frac{k}{2\pi} F_{MN}$. If we also notice that $\epsilon_{\mu\nu\lambda} \epsilon^{\mu\kappa\tau} = - 2\delta_{\nu\lambda}^{\kappa\tau}$ in Lorentzian signature, then we get, with $\mu = -\frac{1}{R}$ but with $\lambda$ in (\ref{mu}) kept as a free parameter,
\bea
-\frac{1}{12} H_{\alpha\beta\gamma}H^{\alpha\beta\gamma} &=& -\frac{1}{4R^2} f_{mn} f^{mn}\cr
-\frac{1}{4} H_{\mu \alpha\beta} H^{\mu \alpha\beta} &=& \frac{1}{4R^2}\(f_{\mu\nu}f^{\mu\nu} - 2 f_{\mu m} f^{\mu m}\)\cr
&& + \frac{\lambda-2}{4R} \epsilon^{\mu\nu\lambda}\omega^{mn} f_{\mu m} f_{\nu\lambda} V_n\cr
&& - \frac{(\lambda-2)^2}{8} \(V^2 f_{\mu m} f^{\mu m} - V^m V_n f_{\mu m} f^{\mu n}\)\cr
\frac{1}{2}\epsilon^{\mu\nu\lambda} \omega^{\alpha\beta\gamma} \partial_{\beta} B_{\mu\alpha} \partial_{\nu} B_{\lambda\gamma} &=& -\frac{1}{2R^2} f_{\mu\nu}f^{\mu\nu} - \frac{\lambda}{4R} \epsilon^{\mu\nu\lambda} \omega^{mn} f_{\mu m} f_{\nu\lambda} V_n
\eea
where all indices $m,n$ are rised by the reduced metric $G_{mn}$. We see that again the nice choice is to take $\lambda = 2$. Dimensional reduction of the measure over the fiber $0 \leq \psi \leq \frac{2\pi}{k}$ gives
\bea
\frac{1}{2\pi}\int d^6 x \sqrt{g} &=& \frac{1}{k} \int d^5 x \sqrt{G} R 
\eea
Finally defining $f_{MN} = \frac{k}{2\pi} F_{MN}$, we get
\bea
S_{e.m.} &=& - \frac{k}{4\pi^2 R} \int d^5 x\sqrt{G}\(\frac{1}{4}\( F_{mn} F^{mn} + 2 F_{\mu m} F^{\mu m} + F_{\mu\nu} F^{\mu\nu}\) + \frac{R}{2} \epsilon_{\mu\nu\lambda} \omega_{mn} F^{\mu m}F^{\nu\lambda} V^n\)
\eea
This we can write in a fully covariant way as
\bea
S_{e.m.} &=& -\frac{1}{4g^2} \int d^5 x \sqrt{G} F_{MN} F^{MN} - \frac{k}{32\pi^2} \int d^5 \sqrt{G} \omega^{MNPQR} V_R F_{MN} F_{PQ}
\eea
with 
\bea
g^2 &=& 4\pi^2 \frac{R}{k}
\eea

\subsubsection{Dimensional reduction of fermionic terms}
Let us reduce the kinetic term for the fermions
\bea
i \bar{\chi} \Gamma^{\alpha} D^{S^3}_{\alpha} \chi
\eea
Recalling that $D_{\alpha}^{S^3} u = 0$ what we have, is 
\bea
i \(\frac{k}{2\pi R}\)^2 \bar{\psi} u^{\dag} \Gamma^{\alpha} u D^{S^3}_{\alpha} \psi 
\eea
Omitting the prefactor, we split this into two terms terms
\bea
\bar{\psi} u^{\dag} \Gamma^m u D^{S^3}_m \psi + \bar{\psi} u^{\dag} \Gamma^{\psi} u D^{S^3}_{\psi} \psi
\eea
which we can compute separately. We first note
\bea
u^{\dag} \Gamma^m u &=& \t \Gamma^m\cr
u^{\dag} \Gamma^{\psi} u &=& \frac{\Gamma_{\psi}}{R^2} \gamma - V^m \t \Gamma_m
\eea
and therefore we have
\bea
 \bar{\psi} \t\Gamma^m D^{S^3}_m \psi + \frac{1}{R} \bar{\psi} \gamma D^{S^3}_{\psi} \psi - V^m \bar{\psi} \t \Gamma_m D^{S^3}_{\psi} \psi
\eea
We insert 
\bea
D_{m}^{S^3} \psi &=& D_m^{S^2} \psi + \frac{1}{2} V_m \t \sigma \psi - \frac{1}{2R} \t \Gamma_m \t \sigma \psi \cr
D_{\psi}^{S^3} \psi &=& \frac{1}{2} \t \sigma \psi
\eea
and we get, after a cancelation of two terms $\sim V_m \bar{\psi} \t \Gamma^m \t \sigma \psi$,
\bea
 \bar{\psi} \t\Gamma^m D^{S^2}_m \psi - \frac{1}{R} \bar{\psi} \t \sigma \psi - \frac{1}{2R} \bar{\psi}\gamma \t\sigma \psi
\eea
However $\bar{\psi} \t \sigma \psi \equiv 0$ due to chiral spinors.

To summarize, dimensional reduction gives the D4 brane action
\bea
\frac{1}{g^2} \int d^5 x \sqrt{G} && \bigg\{ -\frac{1}{4} F_{MN} F^{MN} - \frac{1}{2} G^{MN} \partial_M \Phi \partial_N \Phi + \frac{\Phi}{R} \omega^{mn} F_{mn} - \frac{2}{R^2} \Phi^2\cr
&& - \frac{1}{2} G^{MN} \partial_M \phi^{\hat{i}} \partial_N \phi^{\hat{i}} - \frac{1}{2R^2} \phi^{\hat{i}} \phi^{\hat{i}}\cr
&& + \frac{i}{2} \bar{\psi} \Gamma^M D_M \psi - \frac{i}{4R} \bar{\psi} \t \sigma \(\Gamma_R + \gamma\)\psi \bigg\} - \frac{k}{8\pi^2} \int V \wedge F \wedge F 
\eea
We can make contact with the result in \cite{Nastase:2009zu} by taking
\bea
\mu &=& \frac{2}{R}
\eea
The fact that we shall divide $R$ by $2$ in order to relate with the mass-parameter in ABJM theory through the result in \cite{Nastase:2009zu}, is related with the fact that $\frac{R}{2}$ is the radius of the $S^2$ base-manifold.

\section{Uniqueness}
We expect the D4 brane theory on $\mb{R}^{1,2} \times S^2$ to be rather unique. To establish this, let us make a general ansatz for the Lagrangian,
\bea
\L &=& -\frac{1}{4}F_{MN}F^{MN} + \lambda \omega^{MNPQR} V_M F_{NP} F_{QR}\cr
&& - \frac{1}{2}\partial_M \phi^A \partial^M \phi^A +\frac{a}{R} \Phi \omega^{mn} F_{mn} + \frac{b}{R^2} \phi^2 + \frac{c}{R^2} \phi^{\hat{i}}\phi^{\hat{i}} \cr
&& + \frac{i}{2} \bar{\psi} \Gamma^M D_M \psi + \frac{id}{R} \bar{\psi} \t \sigma \Gamma_R \psi + \frac{i \t d}{R} \bar{\psi} \t \sigma \gamma \psi
\eea
and for the supersymmetry variations
\bea
\delta \phi^A &=& i\bar{\epsilon}\Gamma^A\cr
\delta \psi &=& \frac{1}{2}\Gamma^{MN}\epsilon F_{MN} + \Gamma^M \Gamma_A \epsilon \partial_M \phi^A + \frac{e}{R} \t \sigma \epsilon \phi + \frac{f}{R} \t \sigma \Gamma_R \Gamma_{\hat{i}} \epsilon \phi^{\hat{i}}\cr
\delta A_M &=& i\bar{\epsilon}\Gamma_M \psi
\eea
and assume a Killing spinor equation 
\bea
D_m \epsilon &=& \frac{g}{R} \Gamma_m \t \sigma \Gamma_R \epsilon
\eea
for the supersymmetry parameter. 

In order to kill any components in the supersymmetry variation of $\L$ proportional to mixed components $F_{\mu m}$, we must take
\bea
\gamma \epsilon &=& h \Gamma_R \epsilon\cr
h \t d &=& -d
\eea
and then we get
\bea
\delta \L &&\frac{i \partial_m \phi}{R} \omega^{mn} \bar{\psi} \Gamma_n \epsilon \(2a + e\)\cr
&+&\frac{i \partial_{\mu} \phi}{R} \bar{\psi} \Gamma^{\mu} \t \sigma \epsilon \(e + 2g\)\cr
&+&\frac{i\partial_m\phi^{\hat{i}}}{R} \omega^{mn} \bar{\psi}\Gamma_n \Gamma_{\hat{i}} \Gamma_R \epsilon \(f + 4d\)\cr
&+&\frac{i\partial_{\mu}\phi^{\hat{i}}}{R} \bar{\psi} \Gamma^{\mu} \t\sigma \Gamma_R \Gamma_{\hat{i}}\epsilon \(f-2g-4d\)\cr
&+&\frac{i}{R}\bar{\psi}\Gamma_R\epsilon \omega^{mn}F_{mn} \(g-a)\)\cr
&+&\frac{i}{R}\bar{\psi}\Gamma^{\mu\nu}\t\sigma\Gamma_R\epsilon F_{\mu\nu}\(g + \frac{8}{R}\lambda\)\cr
&+&\frac{i\phi}{R^2} \bar{\psi}\Gamma_R\epsilon \(2eg - 2b\)\cr
&+&\frac{i\phi^{\hat{i}}}{R}\bar{\psi}\Gamma_{\hat{i}}\epsilon \(-2fg-2c-4df\)
\eea
Demanding this variation vanish, we get the solution
\bea
d &=& -\frac{g}{4}\cr
f &=& g\cr
e &=& -2g\cr
a &=& g\cr
b &=& -2g^2\cr
c &=& -\frac{g^2}{2}\cr
\lambda &=& -\frac{g}{8}R
\eea
On $S^2$ with radius $\frac{R}{2}$ we must take $g = \pm 1$. For notational convenience, let us put $g=1$ and absorb the sign by redefining $R$ which we thus shall allow to take both negative and positive values. Furthermore we must have $h = \pm 1$, but we may put $h = 1$ and absorb the sign into the definition of $\gamma$. We then end up with the Lagrangian
\bea
\L &=& -\frac{1}{4}F_{MN}F^{MN} - \frac{R}{8} \omega^{MNPQR} V_M F_{NP} F_{QR}\cr
&& - \frac{1}{2}\partial_M \phi^A \partial^M \phi^A +\frac{1}{R} \Phi \omega^{mn} F_{mn} - \frac{2}{R^2}\phi^2 - \frac{1}{2R^2} \phi^{\hat{i}}\phi^{\hat{i}} \cr
&& + \frac{i}{2} \bar{\psi} \Gamma^M D_M \psi - \frac{i}{4R} \bar{\psi} \t \sigma \Gamma_R \psi + \frac{i}{4R} \bar{\psi} \t \sigma \gamma \psi
\eea
which is invariant under the supersymmetry variation
\bea
\delta \phi^A &=& i\bar{\epsilon}\Gamma^A\cr
\delta \psi &=& \frac{1}{2}\Gamma^{MN}\epsilon F_{MN} + \Gamma^M \Gamma_A \epsilon \partial_M \phi^A - \frac{2}{R} \t \sigma \epsilon \phi + \frac{1}{R} \t \sigma \Gamma_R \Gamma_{\hat{i}} \epsilon \phi^{\hat{i}}\cr
\delta A_M &=& i\bar{\epsilon}\Gamma_M \psi
\eea
where the supersymmetry parameter satisfies the Killing spinor equation 
\bea
D_m \epsilon &=& \frac{1}{R} \Gamma_m \t \sigma \Gamma_R \epsilon
\eea
and the chirality condition
\bea
\gamma \epsilon &=& \Gamma_R \epsilon
\eea
The Lagrangian agrees with what we got by dimensional reduction of M5 brane on the Hopf fiber but we have now also seen that it is also uniquely determined by supersymmetry (up to some conventions that we discussed above) up to an overall normalization constant $\frac{1}{g^2}$, which we fix by other means to be $g^2 = 4\pi^2 \left|\frac{R}{k}\right|$. 

As it may come as a small surprise that the graviphoton term is not a supersymmetry invariant in this situation, let us explicitly show how to compute its supersymmetry variation . We first rewrite it as
\bea
\epsilon^{MNPQR} V_M F_{NP} F_{QR} &=& - \epsilon^{MNPQR} W_{MN} A_P F_{QR}
\eea
where we throw away a total derivative. We then get its variation as
\bea
-2 \epsilon^{MNPQR} W_{MN} F_{QR} \delta A_P
\eea
and specializing to the case that $W_{mn}$ are only nonvanishing components, we get
\bea
-2 \omega^{mn} \epsilon^{\mu\nu\lambda} W_{mn} F_{\mu\nu} \delta A_{\lambda}
\eea
For our case, we have $W_{mn} = -\frac{2}{R^2} \omega_{mn}$ and we make a supersymmetry variation of $A_{\lambda}$, and we get the above variation as
\bea
\frac{8i}{R^2} F_{\mu\nu} \bar{\psi} \Gamma^{\mu\nu} \t \Gamma \epsilon
\eea

\section{The M5 brane coupling constant}\label{coupling}
Supersymmetry determines the form of the M5 brane Lagrangian, but is not sufficient to pin down the value of the M5 brane coupling constant. To determine the coupling constant we must turn to the quantum theory of the M5 brane \cite{Henningson:2004dh}. But it would be nice if we could also determine the coupling constant directly from BLG theory. In \cite{Gustavsson:2010ep} it is shown that if we define the tangenial components of the fluctuation fields in BLG theory according to 
\bea
Y^{\alpha} &=& \frac{\lambda}{2} \omega^{\alpha\beta\gamma} B_{\beta\gamma}\cr
\lambda &=& \frac{\pi R^3}{Nk}
\eea
then we have 
\ben
\int_{S^3} H &=& 2\pi Nk \label{a}
\een
Our interpretation is that $N$ corresponds to the number of M2 branes which are dissolved into the M5 brane. We attribute the presence of the Chern-Simons level $k$ in this formula, to the fact that the M5 brane worldvolume is really the orbifold $S^3/{\mb{Z}_k}$, so that
\bea
\int_{S^3/{\mb{Z}_K}} H &=& 2\pi N
\eea
For the sake of clarity, let us omit transverse scalar fluctuation components. It will be sufficient to consider the tangential components $B_{\alpha\beta}$. Following \cite{Gustavsson:2010ep}, we normalize the BLG Lagrangian as  
\bea
\L &=& -\frac{kN}{2\pi \hbar} \(\frac{1}{2} \<D_{\mu}X^I,D^{\mu}X^I\> + \frac{1}{12} \<\{X^I,X^J,X^K\},\{X^I,X^J,X^K\}\> + ...\)
\eea
where the inner product is unit normalized, and is given by
\bea
\<\bullet\> &=& \frac{1}{N}\tr
\eea
in a matrix realization. For the three-sphere we have \cite{Gustavsson:2010ep}
\bea
\hbar &=& -\frac{R^3}{2N} + \O\(\frac{1}{N^2}\)
\eea
and we match the unit normalized inner product on matrix space to the unit normalized inner product on function space on $S^2\subset S^3$ according to
\bea
\<\bullet\> &=& \frac{1}{\pi R^2}\int d\theta d\varphi \(\frac{R}{2}\)^2 \sin \theta\cr
&=& \frac{1}{\pi R^3} \frac{k}{2\pi} \int_0^{\frac{2\pi}{k}} d\psi \int d\theta d\varphi \frac{R^3}{4}\sin \theta\cr
&=& \frac{k}{2\pi^2 R^3} \int_{S^3/{\mb{Z}_k}} d^3 \sigma \sqrt{g}
\eea
We then get
\bea
\L &=& -\frac{1}{4\pi} \int_{S^3/{\mb{Z}_k}} d\Omega_3 \frac{1}{2}H_{\mu\alpha\beta}H^{\mu\alpha\beta}+...
\eea
and here we can read off the overall constant of the M5 brane Lagrangian, and it agrees with the M5 brane coupling constant in \cite{Henningson:2004dh}.

\section{Summary}\label{summary}
\begin{itemize}
\item
The M5 brane action on $\mb{R}^{1,2} \times S^3$ reads
\bea
S &=& \frac{1}{2\pi} \int d^6 x \Big(-\frac{1}{12} H_{\alpha\beta\gamma}H^{\alpha\beta\gamma} - \frac{1}{4} H_{\mu\alpha\beta} H^{\mu\alpha\beta} + \frac{1}{2} \epsilon^{\mu\nu\lambda} \omega^{\alpha\beta\gamma} \partial_{\beta} B_{\mu\alpha} \partial_{\nu}B_{\lambda\gamma}\cr
&& - \frac{1}{2} g^{\alpha\beta} \partial_{\alpha} Y^A \partial_{\beta} Y^A - \frac{1}{2} \partial_{\mu} Y^A \partial^{\mu} Y^A\cr
&& - \frac{1}{3R} Y \omega^{\alpha\beta\gamma}H_{\alpha\beta\gamma}  - \frac{2}{R^2} Y^2 - \frac{1}{2R^2} Y^{\hat{i}}Y^{\hat{i}}\cr
&& +\frac{i}{2}\(\bar{\chi}\Gamma^{\mu}\partial_{\mu}\chi + \bar{\chi} \Gamma^{\alpha} D_{\alpha}^{(S^3)} \chi\) + \frac{i}{4R} \bar{\chi} \Sigma \Gamma_R \chi \Big)
\eea
It is invariant under the following $(2,0)$ supersymmetry variations 
\bea
\delta Y_A &=& i \bar{\omega} \Gamma_A \chi\cr
\delta B_{\mu\alpha} &=& i \bar{\omega} \Gamma_{\mu}\Gamma_{\alpha} \chi\cr
\delta B_{\alpha\beta} &=& i \bar{\omega} \Gamma_{\alpha\beta} \chi\cr
\delta \chi &=& \frac{1}{2} \Gamma^{\mu}\Gamma^{\alpha\beta} \omega H_{\mu\alpha\beta} + \frac{1}{6} \Gamma^{\alpha\beta\gamma}\omega H_{\alpha\beta\gamma}\cr
&&+\Gamma^{\mu}\Gamma^A \omega \partial_{\mu} Y_A + \Gamma^{\alpha}\Gamma^A \omega \partial_{\alpha} Y_A\cr
&&-\frac{2}{R} \Sigma \omega Y + \frac{1}{R} \Sigma \Gamma_R \Gamma_{\hat{i}} \omega Y^{\hat{i}}
\eea
where the supersymmetry parameter is subject to the chirality condition
\bea
\t \Gamma \Sigma \omega &=& -\omega
\eea
and the Killing spinor equation
\bea
D_{\alpha} \omega &=& -\frac{1}{2R}\Gamma_{\alpha} \Sigma \Gamma_R \omega
\eea
The spinor $\chi$ has the opposite chirality
\bea
\t \Gamma \Sigma \chi &=& \chi
\eea

\item
The corresponding D4 brane action on $\mb{R}^{1,2}\times S^2$, which is obtained by dimensional reduction along the Hopf fiber, is given by 
\bea
\frac{1}{g^2} \int d^5 x \sqrt{G} && \bigg\{ -\frac{1}{4} F_{MN} F^{MN} - \frac{1}{2} G^{MN} \partial_M \Phi \partial_N \Phi + \frac{\Phi}{R} \omega^{mn} F_{mn} - \frac{2}{R^2} \Phi^2\cr
&& - \frac{1}{2} G^{MN} \partial_M \phi^{\hat{i}} \partial_N \phi^{\hat{i}} - \frac{1}{2R^2} \phi^{\hat{i}} \phi^{\hat{i}}\cr
&& + \frac{i}{2} \bar{\psi} \Gamma^M D_M \psi - \frac{i}{4R} \bar{\psi} \t \sigma \(\Gamma_R + \gamma\)\psi \bigg\} -\frac{k}{8\pi^2} \int V \wedge F \wedge F 
\eea
where
\bea
g^2 &=& 4\pi^2 \left| \frac{R}{k}\right|
\eea
The action is invariant under the supersymmetry variations
\bea
\delta \phi_A &=& i \bar{\epsilon} \Gamma_A \psi\cr
\delta A_M &=& i \bar{\epsilon} \Gamma_M \psi\cr
\delta \psi &=& \frac{1}{2} \Gamma^{MN} \epsilon F_{MN} + \Gamma^M \Gamma^A \epsilon \partial_M \phi_A\cr
&& +\frac{2}{R} \t \sigma \epsilon \Phi + \frac{1}{R} \t \sigma \Gamma_R \Gamma_{\hat{i}} \epsilon \phi^{\hat{i}}
\eea
where the spinor and the supersymmetry parameter have the same chirality
\bea
\psi &=& \frac{\Gamma_{\psi}}{R} \psi\cr
\epsilon &=& \frac{\Gamma_{\psi}}{R} \epsilon
\eea
and where the supersymmetry parameter is further restricted by an additional Weyl projection
\bea
(\Gamma_R + \gamma)\epsilon &=& 0
\eea
and satisfies the Killing spinor equation
\bea
D_{m}^{S^2} \epsilon &=& \frac{1}{R}\t \Gamma_m \t \sigma \Gamma_R \epsilon
\eea
associated to $S^2$ base-manifold of radius $\frac{R}{2}$.

In the limit $R \rightarrow \infty$ we just have the usual flat space sYM variations for a constant supersymmetry parameter, and then supersymmetry obviously enhances from $8$ to $16$ supersymmetries. 
\end{itemize}

\section{Open questions}
Let us end this paper by listing some open questions,
\begin{itemize}
\item How can we deconstruct the graviphoton term from ABJM? One appealing idea seems to be that one should really take the gauge group as $U(N)\times U(N-1)$ rather than $U(N)\times U(N)$. This might be justfied by that the GRVV matrices generate the algebra of $N \times (N-1)$ matrices. In that case Higgsing amounts to an additional $U(1)$ CS-term which is similar, if not identical, to the graviphoton term in five dimensions upon an integration by parts.

Let us consider a compact euclidean spacetime $T^3\times S^2$ and assume the graviphoton term of the form (with $W = dV$ and $\int_{S^2} W = 2\pi$)
\bea
\frac{k}{8\pi^2}\int_{(x,\sigma) \in T^3 \times S^2} W(\sigma)\wedge A(x,\sigma) \wedge dA(x,\sigma)
\eea
If then we let $B_4$ have $T^3$ as boundary, we have
\bea
\frac{k}{8\pi^2} \int_{S^2} W(\sigma) \int_{B_4} F(x,\sigma) \wedge F(x,\sigma)
\eea
and now the integral over $B_4$ is integer quantized, hence it can not depend smoothly on $\sigma\in S^2$. Then we can perform the integral over $S^2$, and get 
\bea
\frac{k}{4\pi}\int_{T^3} A(x) \wedge dA(x)
\eea
which is the CS term on $T^3$.

\item Why does our dimensional reduction to the D4 brane action appear to break half the supersymmetry? We know that ABJM has $12$ supercharges, which are enhanced to $16$ supercharges when $k=1,2$. By deconstruction, we would expect to find at least $12$ supercharges on the D4 brane. 

\item It is true that dimensional reduction may be best motivated while taking $k\rightarrow \infty$. However, recently it was conjectured that D4 and M5 are dual, and such a duality should be valid for any size of the compactification radius. Then we should be able to take $k=1$ and relate D4 and M5 as dual theories. Should we then expect the D4 to have $16$ supersymmetries, whereof half of these supersymmetries are hidden?

\end{itemize}

\subsubsection*{Acknowledgements}
I would like to thank Soo-Jong Rey, Takao Suyama, Bengt E W Nilsson for discussions. This work was supported by NRF Mid-career Researcher Program 2011-0013228.

\newpage

\appendix

\section{Riemann geometry in vielbein formulation}\label{A}
The reason we are interested in vielbein formulation is because it is needed to define the covariant derivative of a spinor field. For a vector field it is not needed, but nevertheless it can be used also for vector fields. So by using the vielbein formulation we have a unified treatment of both spinor and vector fields, which we find very nice. 

\subsection{Covariant derivative}

Let us denote a generic tensor-spinor field as $\psi$, and let us refer to two different local coordinate systems by coordinates $x^{\mu}$ and $q^{\alpha}$ respectively. Let us make the following ansatz for the covariant derivative
\bea
D_{\mu} \psi &=& \partial_{\mu} \psi + \omega_{\mu} \psi
\eea
in one of these coordinate systems. If we assume that under a change of coordinates, the spinor-tensor transforms as
\bea
\psi(q) &=& g(q,x) \psi(x)
\eea
where the transition matrix $g$ will be specified shortly, then the covariant derivative shall transform covariantly as
\bea
D_{\alpha} \psi(q) &=& g(q,x) \frac{\partial x^{\mu}}{\partial q^{\alpha}} D_{\mu} \psi(x)
\eea
These conditions are solved by assuming the transformation law
\ben
\omega_{\mu} &=& \frac{\partial q^{\alpha}}{\partial x^{\mu}} \(g^{-1} \Omega_{\alpha} g + g^{-1} \partial_{\alpha} g\)\label{conntransf}
\een
which shows that $\omega_{\mu}$ shall transform like a connection one-form.

Let us now specify the transition matrix $g$ as we go from one coordinate system, $x^{\mu}$, to another one, $q^{\alpha}$. For a vector field, we have 
\bea
\psi^{\alpha}(q) &=& \frac{ \partial q^{\alpha} }{ \partial x^{\mu} } \psi^{\mu}(x)
\eea
Now let us introduce vielbeins $e^i_{\mu}(x)$ and $f^i_{\alpha}(q)$ corresponding to these two coordinate systems. Then we define 
\bea
\psi^i(q) &=& f^i_{\alpha} \psi^{\alpha}(q)\cr
\psi^i(x) &=& e^i_{\mu} \psi^{\mu}(x)
\eea
and these are now related as
\bea
\psi^i(q) &=& g^i{}_j(q,x) \psi^j(x)
\eea
where the transition matrix, and its inverse, are given by
\bea
g^i{}_j(q,x) &=& f^i_{\alpha} \frac{\partial q^{\alpha}}{\partial x^{\mu}} e^{\mu}_j\cr
{g^{-1}}^i{}_j(x,q) &=& e^i_{\mu}\frac{\partial x^{\mu}}{\partial q^{\alpha}} f^{\alpha}_j
\eea

We will now show that we may construct such a connection one-form $\omega$ explicitly as 
\ben
\omega_{\mu j}^i &=& e_{\lambda}^i \bar{D}_{\mu} e_j^{\lambda}\label{conn}
\een
Here the covariant derivative $\bar{D}_{\mu}$ with a bar is blind to flat indices $i,j,...$. Thus 
\bea
\bar{D}_{\mu} e_j^{\lambda} &:=& \partial_{\mu} e_j^{\lambda} + \Gamma^{\lambda}_{\mu\nu} e^{\nu}_j
\eea
Contracting (\ref{conn}) by $e^{\rho}_i$ we get
\bea
\bar{D}_{\mu} e_j^{\rho} - \omega_{\mu j}^i e_i^{\rho} &=& 0
\eea
which we can express as 
\bea
D_{\mu} e_j^{\rho} &=& 0
\eea
for the full covariant derivative which acts on both curved and flat indices.

Assuming that $\omega$ transforms like a connection one-form (\ref{conntransf}) we can derive the correct transformation rule for the Christoffel symbol. Let us temporarily define the non-covariant nonsensical object (for the sake of notational simplicity only)
\bea
\omega_{\mu\nu}^{\lambda} &=& \omega_{\mu j}^i e^j_{\nu} e_i^{\mu} 
\eea
Then the Christoffel symbols are given by
\bea
\Gamma_{\mu\nu}^{\lambda} &=& \omega_{\mu\nu}^{\lambda} - e^j_{\nu} \partial_{\mu} e_j^{\lambda}\cr
\Gamma_{\alpha\beta}^{\gamma} &=& \omega_{\alpha\beta}^{\gamma} - f^j_{\beta} \partial_{\alpha} e^{\gamma}_j
\eea
By inserting explicit expressions for $g$ and $g^{-1}$ in (\ref{conntransf}), we derive the transformation rule of the Christoffel symbol
\bea
\Gamma_{\mu\nu}^{\lambda} &=& \frac{\partial q^{\alpha}}{\partial x^{\mu}} \frac{\partial q^{\beta}}{\partial x^{\nu}} \frac{\partial x^{\lambda}}{\partial q^{\gamma}} \Gamma^{\gamma}_{\alpha\beta} + \frac{\partial x^{\lambda}}{\partial q^{\alpha}} \frac{\partial q^{\alpha}}{\partial{x^{\mu}}\partial x^{\nu}}
\eea
Each step in this computation can be revered, so that by assuming this transformation for the Christoffel symbol we can derive the transformation rule (\ref{conntransf}).

Let us assume the tangent space group $SO(d)$ with Lie algebra 
\bea
[M_{ij},M_{kl}] &=& -4 \delta_{[i}^{[k} M_{j]}{}^{l]}
\eea
In the vector representation 
\bea
(M_{kl})^{ij} &=& 2\delta_{kl}^{ij}
\eea
and in the spinor representation
\bea
M_{kl} &=& \frac{1}{2} \Gamma_{kl}
\eea
We may then simply replace 
\bea
\omega^{ij} &=& \frac{1}{2} \omega^{kl} (M_{kl})^{ij}
\eea
in the vector representation, with an arbitrary representation of $SO(d)$,
\bea
\omega &=& \frac{1}{2} \omega^{kl} M_{kl}
\eea
and this is how we obtain the covariant derivative on acting on a generic spinor-tensor.  In particular we get for a spinor,
\bea
D_{\alpha} &=& \partial_{\alpha} + \omega_{\alpha}\cr
\omega_{\alpha} &=& \frac{1}{4}\omega_{\alpha}^{ij}\Gamma_{ij}
\eea
The gamma matrices $\Gamma_i$ are taken as some given fixed constant matrices. They are gauge invariant,
\bea
\Gamma_i &=& g_{ij} g \Gamma_j g^{-1}
\eea
where\footnote{In our convention, $M_{ij}$ is antihermitian so there is no factor of $i$ in the exponent.} 
\bea
g &=& e^{\frac{1}{2}\Lambda^{ij}M_{ij}}
\eea
in the vector (where it is written $g_{ij}$) and the spinor representation respectively. 
Infinitesimally we have
\bea
\Lambda_{ij} \Gamma_j + [\Lambda,\Gamma_i] &=& 0
\eea
This is a consequence of $SO(d)$ Clifford algebra
\bea
\{\Gamma_i,\Gamma_j\} &=& 2\delta^{ij}
\eea
from which we can derive the identity $[\Gamma{ij},\Gamma_k] = - 4 \delta_{k[i}\Gamma_{j]}$. It follows that the gamma matrices are covariantly constant
\bea
D_{\mu} \Gamma_i &=& 0
\eea
Obviously the $\Gamma_i$, satisfying the $SO(d)$ Clifford algebra, may be chosen as constant matrices of some standard form. The covariant derivative thus reduces to
\bea
D_{\mu} \Gamma_i &=& \Omega_{\mu}^{ij} \Gamma_j + [\Omega_{\mu},\Gamma_i]
\eea
But the right-hand side is is nothing but the invariance condition of gamma matrices so it vanishes.

We convert to curved space by a vielbein
\bea
\Gamma_{\mu} &=& e_{\mu}^i \Gamma_i
\eea
and these are again covariantly constant
\bea
D_{\mu} \Gamma_{\nu} &=& 0
\eea
due to the fact that 
\bea
D_{\mu} e_{\nu}^i &=& 0
\eea

\subsection{Lie derivative}\label{Lie}
A first attempt is to define the Lie derivative of any object as
\bea
\L_v V(x) &=& V'(x) - V(x)
\eea
where $v^{\mu}(x)$ initially is an arbitrary infinitesimal displacement vector field, and $V'$ denotes the transformed quantity. However if $V$ is a spinor fields, it appears we must in addition assume that $v^{\mu}$ is a Killing vector field. Let us anyway first compute the Lie derivative explicitly for an ordinary tensor field $V_{\mu_1 \mu_2 ...}(x)$. Then we have
\bea
\L_v V_{\mu_1\mu_2...}(x) &=& v^{\nu} \partial_{\nu} V_{\mu_1\mu_2...}(x) + \partial_{\mu} v^{\nu} T_{\nu\mu_1...} + \partial_{\mu} v^{\nu} T_{\mu_1 \nu ...} + ...
\eea
We can substitute ordinary derivatives with covariant ones for free. As it turns out, all the Christoffel connections cancel. So we have
\bea
\L_v V_{\mu_1\mu_2...}(x) &=& v^{\nu} D_{\nu} V_{\mu_1\mu_2...}(x) + D_{\mu_1} v^{\nu} T_{\nu\mu_1...} + D_{\mu_2} v^{\nu} T_{\mu_1 \nu ...} + ...
\eea
If $v$ is a Killing vector field, then $\L_v g_{\mu\nu} = 0$ by definition, where $g_{\mu\nu}$ denotes the metric tensor. We have some freedom to choose a vielbein associated to the metric, since we can rotate the local rotation group index. Let us for simplicity assume that we make this choice so that also $\L_v e_{\mu}^i = 0$. It just means that we use the same vielbein to the transformed metric, which since $v$ is a Killing vector, is the same as the original metric, $g'_{\mu\nu}(x) = g_{\mu\nu}(x)$. Then, by using the Leibniz rule for the Lie derivative, we can express the above Lie derivative in tangent space indices as
\bea
\L_v V_{i_1 iu_2...}(x) &=& v^{\nu} D_{\nu} V_{i_1 i_2...}(x) + D_{i_1} v^{j} T_{j i_1...} + D_{i_2} v^{j} T_{i_1 j ...} + ...
\eea
If we further write this in terms of rotation group generators $M_{ij}$, we have
\ben
\L_v V(x) &=& v^{\nu} D_{\nu} V + \frac{1}{2} D_k v_l M^{kl} V\label{Lieder}
\een
which is now a completely general formula for the Lie derivative, which applies to tensor and spinors alike.

Such a general Lie derivative (it was called Lorentz-Lie derivative) has been introduced also in \cite{Ortin:2002qb}.

\newpage

\section{Three-sphere in vielbein formulation}\label{B}
In ${\mb{R}}^4$ we define the relation between Cartesian and Polar coordinates as 
\bea
x^1 &=& R \sin \theta \sin \varphi \sin \psi\cr
x^2 &=& R \sin \theta \sin \varphi \cos \psi\cr
x^3 &=& R \sin \theta \cos \varphi\cr
x^4 &=& R \cos \theta
\eea
The metric is
\bea
ds^2 &=& \sum_{i=1}^4 dx^i dx^i \cr
&=& R^2 \(d\theta^2 + \sin^2 \theta d\varphi^2 + \sin^2 \theta \sin^2 \varphi d\psi^2\) + dR^2
\eea
We have the diagonal vielbein
\bea
e^{i}_{\mu} &=& \(\begin{array}{cccc}
R\sin \theta \sin \varphi & 0 & 0 & 0\\
0 & R\sin \theta & 0 & 0\\
0 & 0 & R & 0\\
0 & 0 & 0 & 1
\end{array}\)
\eea
where rows are associated with $q^{\mu} = (\psi, \varphi, \theta, R)$ components in that order. We may also consider another vielbein
\bea
E^i_{\mu} &=& \frac{\partial x^i}{\partial q^{\mu}}
\eea
which is not diagonal.

The spin connection is most easily computed from the torsion free condition. Let us denote the spin connections associated to these two choices of vielbeina as $\omega$ and $\Omega$ respectively. Defining the one-forms $e^i = e^i_{\mu} dq^{\mu}$ and $E^i = E^i_{\mu} dq^{\mu}$ respectively, the torsion free conditions read
\bea
0 &=& de^i + \omega^{ij} \wedge e^j\cr
0 &=& dE^i + \Omega^{ij} \wedge E^j  
\eea
The two choices of vielbein is related by a gauge transformation
\bea
E^i_{\mu} &=& g^i{}_j e^j_{\mu}\cr
g^i{}_j &=& E^i_{\mu} e^{\mu}_j
\eea
and
\bea
\omega^{ij} &=& (g^{-1} dg)^{ij} + (g^{-1} \Omega g)^{ij}
\eea
Let us refer to the matrix $g^i{}_j$ as the transition matrix.

The covariant derivative is
\bea
D_{\mu}\psi(e) &=& \partial_{\mu}\psi(e) + \omega_{\mu}\psi(e)\cr
D_{\mu}\psi(E) &=& \partial_{\mu}\psi(E) + \Omega_{\mu}\psi(E)
\eea
Under the change of vielbein, a spinor transforms as
\bea
\psi(E) &=& g \psi(e)
\eea
Gamma matrices shall transform so that $\Gamma_i \psi$ transforms like a spinor with one flat index 
\bea
\Gamma_i(E) \psi(E) &=& g \Gamma_j(e)g^j{}_i \psi(e)
\eea
It implies that
\bea
\Gamma_i(E) &=& g \Gamma_i(e) g^{-1} g^j{}_i
\eea
but since we have the invariance condition of gamma matrices, this amounts to 
\bea
\Gamma_i(E) &=& \Gamma_i(e)
\eea
Since $E^i = dx^i$ is an exact differential form, it is closed, 
\bea
dE^i &=& 0
\eea
and the torsion free condition gives 
\bea
\Omega^{ij} &=& 0
\eea
and then 
\bea
\omega^{ij} &=& (g^{-1}dg)^{ij}
\eea
which indeed is a flat connection. 

We can also derive this in a more direct way. As $\Omega_{\alpha} = 0$, we clearly have
\bea
D_{\alpha} \psi(E) = \partial_{\alpha} \psi(E)
\eea
If we then write $\psi(E) = g \psi(e)$ we get
\bea
D_{\alpha} \psi(E) &=& g D_{\alpha} \psi(e)\cr
D_{\alpha} \psi(e) &=& \partial_{\alpha} \psi(e) + g^{-1} \partial_{\alpha} g \psi(e)
\eea
We again get
\bea
\omega_{\alpha} &=& g^{-1} \partial_{\alpha} g
\eea

One question we would now like to answer is how to extract the spin-connection on $S^3$ from the spin connection on ${\mb{R}}^4$. To answer this question, it is most convenient to work with the diagonal vielbein. We compute the spin-connection $\omega$ from the torsion free condition, with the result
\ben
\omega^{23} &=& \cos \theta d\varphi\cr
\omega^{13} &=& \cos \theta \sin \varphi d\psi\cr
\omega^{12} &=& \cos \varphi d\psi\cr
\omega^{14} &=& \sin \theta \sin \varphi d\psi\cr
\omega^{24} &=& \sin \theta d\varphi\cr
\omega^{34} &=& d\theta\label{spinconnection}
\een
In the diagonal vielbein basis it is clear that the components $i,j\neq 4$ are associated with $S^3$. The components which do involve index $4$ may also be expressed as
\bea
\omega^{i4} &=& \frac{1}{R} e^i
\eea
for $i=1,2,3$. Also we may note that for all $i,j$,
\bea
\omega_R^{ij} &=& 0
\eea

But as we explained above we could also obtain $\omega$ from the transitition matrix $g^i{}_j = E^i_{\mu} e^{\mu}_j$. This matrix can be computed explicitly with the result
\bea
g &=& \(\begin{array}{cccc}
\cos \psi & \cos \varphi \sin \psi & \cos \theta \sin \varphi \sin \psi & \sin \theta \sin \varphi \sin \psi\\
-\sin \psi & \cos \varphi \cos \psi & \cos \theta \sin\varphi \cos \psi & \sin \theta \sin \varphi \cos \psi\\
0 & -\sin \varphi & \cos \theta \cos \varphi & \sin \theta \cos \varphi\\
0 & 0 & -\sin \theta & \cos \theta
\end{array}\) \cr
&=& \(\begin{array}{cccc}
\cos \psi & \sin \psi & 0 & 0\\
-\sin \psi & \cos \psi & 0 & 0\\
0 & 0 & 1 & 0\\
0 & 0 & 0 & 1
\end{array}\)
\(\begin{array}{cccc}
1 & 0 & 0 & 0\\
0 & \cos \varphi & \sin \varphi & 0 \\
0 & -\sin \varphi & \cos \varphi & 0\\
0 & 0 & 0 & 1
\end{array}\)
\(\begin{array}{cccc}
1 & 0 & 0 & 0\\
0 & 1 & 0 & 0\\
0 & 0 & \cos \theta & \sin \theta\\
0 & 0 & -\sin \theta & \cos \theta
\end{array}\)
\eea
We can write this as 
\bea
g^i{}_{j} &=& \(e^{\psi M_{12}}\)^i{}_{k} \(e^{\varphi M_{23}}\)^{k}{}_l \(e^{\theta M_{34}}\)^l{}_{j}
\eea
in the vector representation
\bea
(M_{ij})_{kl} &=& 2\delta_{ij}^{kl}
\eea
In the spinor representation
\bea
M_{ij} &=& \frac{1}{2} \Gamma_{ij}
\eea
and
\bea
g &=& e^{\frac{1}{2}\psi \Gamma_{12}} e^{\frac{1}{2}\varphi \Gamma_{23}} e^{\frac{1}{2}\theta \Gamma_{34}}
\eea
We may now compute the spin connection as
\bea
g^{-1}\partial_{\theta} g &=& \frac{1}{2}\Gamma_{34}\cr
g^{-1}\partial_{\varphi}g &=& \frac{1}{2}\Gamma_{23} \cos \theta + \frac{1}{2} \Gamma_{24} \sin \theta\cr
g^{-1}\partial_{\psi}g &=& \frac{1}{2} \(\Gamma_{13} \cos \theta \sin \varphi + \Gamma_{12} \cos \varphi\) + \frac{1}{2} \Gamma_{14} \sin \theta\sin \varphi\cr
g^{-1} \partial_R g &=& 0
\eea
which indeed agrees with the result (\ref{spinconnection}) if we recall that in the spinor representation
\bea
\omega_{\alpha} &=& \frac{1}{4} \omega_{\alpha}^{ij} \Gamma_{ij}
\eea

\subsection{Killing spinor equation on $S^3$}\label{killing}
We assume a constant spinor $\epsilon$ on ${\mb{R}}^4$,
\ben
\partial_i \epsilon(x) &=& 0\label{constantspinor}
\een
We then transform to polar coordinates and get in particular that
\bea
\partial_{\alpha} \epsilon(E) &=& 0
\eea
We will not be interested in the $\partial_R$ derivative here. We note that $\epsilon(E) = \epsilon(x)$. We transform to diagonal vielbein basis
\bea
\psi(E) &=& g \psi(e)
\eea
and we get
\bea
\partial_{\alpha} \epsilon(e) + \omega_{\alpha} \epsilon(e) &=& 0
\eea
Now we split 
\bea
\omega_{\alpha} &=& \omega^{(S^3)}_{\alpha} + \frac{1}{2R} e_{\alpha}^i e_R^j \Gamma_{ij}
\eea
We now derive from (\ref{constantspinor}) one of the two possible Killing spinor equations on $S^3$,
\ben
D_{\alpha}^{(S^3)} \epsilon(e) &=& -\frac{1}{2R} \Gamma_{\alpha}\Gamma_R \epsilon(e)\label{killingeq}
\een
where we also introduced 
\bea
\Gamma_{\alpha} &=& e_{\alpha}^i \Gamma_i\cr
\Gamma_R &=& e_R^i \Gamma_i
\eea
The other Killing spinor on $S^3$ reads
\bea
D_{\alpha}^{(S^3)} \epsilon(e) &=& \frac{1}{2R} \Gamma_{\alpha}\Gamma_R \epsilon(e)
\eea
but we do not get this equation from a constant spinor on ${\mb{R}}^4$. 

We note that 
\ben
D_{\alpha}^{(S^3)} \Gamma_{\beta} &=& 0\cr
D_{\alpha}^{{\mb{R}}^4} \Gamma_{\beta} &=& 0\label{cov}
\een
It is clear that $\Gamma_{\mu} = e_{\mu}^i \Gamma_i$ is covariantly constant in ${\mb{R}}^4$. But since the vielbein is diagonal, it is also true that $\Gamma _{\alpha} = e_{\alpha}^i \Gamma_i$ are covariantly constant on $S^3$. We can also check this explicitly. We compute
\bea
D_{\alpha}^{{\mb{R}}^4} \Gamma_{\beta} &=& D_{\alpha}^{(S^3)} \Gamma_{\beta} - \Gamma_{\alpha\beta}^R \Gamma_R + \frac{1}{2R} [\Gamma_{\alpha}\Gamma_R, \Gamma_{\beta}]\cr
D_{\alpha}^{{\mb{R}}^4} \Gamma_R &=& D_{\alpha}^{(S^3)} \Gamma_R - \Gamma_{\alpha R}^{\gamma} \Gamma_{\gamma} + \frac{1}{2R} [\Gamma_{\alpha} \Gamma_R,\Gamma_R]
\eea
We then note that 
\bea
 - \Gamma_{\alpha\beta}^R \Gamma_R + \frac{1}{2R} [\Gamma_{\alpha}\Gamma_R, \Gamma_{\beta}] &=& 0\cr
-\Gamma_{\alpha R}^{\gamma} \Gamma_{\gamma} + \frac{1}{2R} [\Gamma_{\alpha} \Gamma_R,\Gamma_R] &=& 0
\eea
by using the fact that the only non-vanishing Christoffel symbols are
\bea
\Gamma_{\alpha\beta}^{\gamma} &&{\mbox{(on $S^3$)}}\cr
\Gamma_{\alpha\beta}^R &=& -\frac{1}{R}g_{\alpha\beta}\cr
\Gamma_{\alpha R}^{\beta} &=& \frac{1}{R} \delta^{\alpha}_{\beta}
\eea
Finally 
\bea
D_{\alpha} e_R^i &=& 0
\eea

\subsection{Curvature of $S^3$}
The curvature two-form on the $S^3$ submanifold in any representation is given by
\bea
R_{\alpha\beta} &=& \partial_{\alpha} \omega_{\beta} - \partial_{\beta} \omega_{\alpha} + [\omega_{\alpha},\omega_{\beta}]
\eea
One may connect the curvature two-form with the Riemann curvature tensor by noting 
\bea
D_{\alpha} \omega_{\beta j}^i &=& \partial_{\alpha} \omega_{\beta j}^i - \Gamma_{\alpha\beta}^{\gamma}\omega_{\gamma j}^i\cr
R_{\alpha\beta}{}^i{}_j &=& D_{\alpha} \omega_{\beta j}^i - D_{\beta} \omega_{\alpha j}^i + [\omega_{\alpha},\omega_{\beta}]^i{}_j\cr
\omega_{\alpha j}^i &=& e_{\gamma}^i D_{\alpha} e^{\gamma}_j
\eea
and thus restrict ourselves to the vector representation (indices $i,j,...$). Then one finds
\bea
R_{\alpha\beta}{}^i{}_j &=& e^i_{\gamma} [D_{\alpha},D_{\beta}] e^{\gamma}_j
\eea
Defining the Riemann curvature tensor through the relation
\bea
[D_{\alpha},D_{\beta}]V^{\gamma} &=& R_{\alpha\beta}{}^{\gamma}{}_{\delta} V^{\delta}
\eea
we thus find
\bea
R_{\alpha\beta}{}^i{}_j &=& R_{\alpha\beta}{}^{\gamma}{}_{\delta} e^i_{\delta} e_j^{\gamma}
\eea

In the case of $S^3$ we get 
\bea
R_{ab} &=& \frac{1}{R^2}e^a \wedge e^b
\eea
which more explicitly means that 
\bea
R_{\alpha\beta ab} &=& \frac{1}{R^2} e^a_{[\alpha} e^b_{\beta]}
\eea
which can be converted to curved indices by contraction by two vielbeins, and we get
\bea
R_{\alpha\beta\gamma\delta} &=& \frac{1}{2R^2} \(g_{\alpha\gamma} g_{\beta\delta} - g_{\alpha\delta} g_{\beta\gamma}\)
\eea
Let us finally check this formula for the $12$ component,
\bea
R_{12} &=& d\omega_{12} + \omega_{13}\wedge \omega_{32}\cr
&=& \sin \varphi \sin^2 \theta d\psi \wedge d\varphi
\eea
Indeed this agrees with $\frac{1}{R^2}e^1 \wedge e^2$.

\newpage
\section{Verification of M5 brane supersymmetry}\label{M5close}
We vary the bosonic terms in the M5 brane Lagrangian (\ref{Lagr}) and we find $10$ terms after an integration by parts
{
\tiny
\bea
&& \frac{i}{2} D_{\alpha} H^{\alpha\beta\gamma} \bar{\omega} \Gamma_{\beta\gamma}\chi + \frac{i}{2} \partial_{\mu} H^{\mu\alpha\beta} \bar{\omega}\Gamma_{\beta\gamma}\chi + i D_{\beta}H^{\mu\alpha\beta} \bar{\omega}\Gamma_{\mu}\Gamma_{\alpha}\chi - i \epsilon^{\mu\nu\lambda}\omega^{\alpha\beta\gamma}\bar{\omega}\Gamma_{\mu}\Gamma_{\alpha}\chi D_{\beta}\partial_{\nu} B_{\lambda\gamma} \cr
&& + i D^M \partial_M Y^A \bar{\omega}\Gamma^A \chi + \frac{i}{3R} \slashed{H} \bar{\omega}\Gamma_R\chi - \frac{2i}{R} \partial_{\alpha}Y \bar{\omega} \Gamma^{\alpha} \Sigma \chi - \frac{4i}{R} Y \bar{\omega} \Gamma_R \chi - \frac{i}{R^2} Y^{\hat{i}} \bar{\omega} \Gamma^{\hat{i}} \chi
\eea}
and we vary the fermionic terms and we find $22$ terms
{
\tiny
\bea
&& \frac{i}{2}\bar{\chi} \Gamma^{\mu} \Gamma^{\nu} \Gamma^{\beta\gamma} \omega \partial_{\mu} H_{\nu\beta\gamma} + \frac{i}{6} \bar{\chi} \Gamma^{\mu} \Gamma^{\beta\gamma\delta} \omega \partial_{\mu} H_{\beta\gamma\delta} + i \bar{\chi} \Gamma^A \omega \partial^{\mu} \partial_{\mu} Y_A - \frac{2i}{R} \bar{\chi} \Gamma^{\mu} \Sigma \omega \partial_{\mu} Y + \frac{i}{R} \bar{\chi} \Gamma^{\mu} \Sigma \Gamma_R \Gamma_{\hat{i}} \omega \partial_{\mu} Y^{\hat{i}}\cr
&&- \frac{i}{2} \bar{\chi} \Gamma^{\nu}\Gamma^{\alpha\beta\gamma}\omega D_{\alpha} H_{\nu\beta\gamma} - i \bar{\chi} \Gamma^{\nu} \Gamma^{\gamma} \omega D^{\alpha} H_{\nu\alpha\gamma} + i \bar{\chi} \Gamma^A\omega D^{\alpha}\partial_{\alpha} Y_A - \frac{2i}{R} \bar{\chi} \Gamma^{\alpha} \Sigma \omega \partial_{\alpha} Y + \frac{i}{R} \bar{\chi} \Gamma^{\alpha} \Sigma \Gamma_R \Gamma_{\hat{i}} \omega \partial_{\alpha} Y_{\hat{i}} + \frac{i}{2} \Gamma^{\gamma\delta} \omega D^{\alpha} H_{\alpha\gamma\delta}\cr
&& -\frac{i}{2} \bar{\chi} \Gamma^{\nu} \Gamma^{\alpha\beta\gamma}\Gamma_{\alpha}M \omega H_{\nu\beta\gamma} - i \bar{\chi} \Gamma^{\nu} \Gamma^{\gamma} \Gamma^{\beta} M \omega H_{\nu\beta\gamma} + \frac{i}{6} \bar{\chi} \Gamma^{\alpha} \Sigma \Gamma_{\alpha} M \omega \slashed{H} + i \bar{\chi} \Gamma^{\beta} \Gamma^A M \omega \partial_{\beta} Y_A - \frac{2i}{R} \bar{\chi} \Gamma^{\alpha} \Sigma \Gamma_{\alpha} M \omega Y + \frac{i}{R} \bar{\chi}\Gamma^{\alpha} \Sigma \Gamma_R \Gamma_{\hat{i}} \Gamma_{\alpha} M \omega Y_{\hat{i}}\cr
&& + \frac{i}{4R} \bar{\chi} \Sigma \Gamma_R \Gamma^{\mu}\Gamma^{\alpha} \omega H_{\mu\alpha\beta} + \frac{i}{12 R} \bar{\chi} \Gamma_R \omega \slashed{H} + \frac{i}{2R} \bar{\chi} \Sigma \Gamma_R \Gamma^M \Gamma^A \omega \partial_M Y^A - \frac{i}{R^2} \bar{\chi} \Gamma_R \omega Y + \frac{i}{2R^2} \bar{\chi} \Gamma_{\hat{i}} \omega Y^{\hat{i}}
\eea}
where we have defined
\bea
D_{\alpha} \omega &=& \Gamma_{\alpha} M \omega
\eea
and 
\bea
\slashed{H} &=& \omega^{\alpha\beta\gamma} H_{\alpha\beta\gamma}
\eea
We then match similar terms. Then at the end of the day we find a complete cancelation of all terms if and only if we take 
\bea
M &=& -\frac{1}{2R} \Sigma \Gamma_R
\eea
This is an independent confirmation of our Killing spinor equation for the supersymmetry parameter.


\newpage


\begin{thebibliography}{999}

\bibitem{Lambert:2010iw}
  N.~Lambert, C.~Papageorgakis, M.~Schmidt-Sommerfeld,
  ``M5-Branes, D4-Branes and Quantum 5D super-Yang-Mills,''
  JHEP {\bf 1101}, 083 (2011).
  [arXiv:1012.2882 [hep-th]].


\bibitem{Douglas:2010iu}
  M.~R.~Douglas,
  ``On D=5 super Yang-Mills theory and (2,0) theory,''
  JHEP {\bf 1102}, 011 (2011).
  [arXiv:1012.2880 [hep-th]].

\bibitem{Verlinde:1995mz}
  E.~P.~Verlinde,
  ``Global aspects of electric - magnetic duality,''
  Nucl.\ Phys.\  {\bf B455}, 211-228 (1995).
  [hep-th/9506011].

\bibitem{Witten:1995gf}
  E.~Witten,
  ``On S duality in Abelian gauge theory,''
  Selecta Math.\  {\bf 1}, 383 (1995).
  [hep-th/9505186].


\bibitem{Ho:2011ni}
  P.~-M.~Ho, K.~-W.~Huang, Y.~Matsuo,
  ``A Non-Abelian Self-Dual Gauge Theory in 5+1 Dimensions,''
  JHEP {\bf 1107}, 021 (2011).
  [arXiv:1104.4040 [hep-th]].


\bibitem{Bolognesi:2011rq}
  S.~Bolognesi, K.~Lee,
  ``1/4 BPS String Junctions and $N^3$ Problem in 6-dim (2,0) Superconformal Theories,''
  
  [arXiv:1105.5073 [hep-th]].

\bibitem{Gustavsson:2009qd}
  A.~Gustavsson,
  ``M5 brane from mass deformed BLG theory,''
  JHEP {\bf 0911}, 071 (2009)
  [arXiv:0909.2518 [hep-th]].


\bibitem{Gustavsson:2010ep}
  A.~Gustavsson,
  ``Five-dimensional super Yang-Mills theory from ABJM theory,''
  JHEP {\bf 1103}, 144 (2011).
  [arXiv:1012.5917 [hep-th]].


\bibitem{Nastase:2009zu}
  H.~Nastase and C.~Papageorgakis,
  ``Fuzzy Killing Spinors and Supersymmetric D4 action on the Fuzzy 2-sphere
  from the ABJM Model,''
  JHEP {\bf 0912}, 049 (2009)
  [arXiv:0908.3263 [hep-th]].


\bibitem{Andrews:2006aw}
  R.~P.~Andrews, N.~Dorey,
  ``Deconstruction of the Maldacena-Nunez compactification,''
  Nucl.\ Phys.\  {\bf B751}, 304-341 (2006).
  [hep-th/0601098].


\bibitem{Abrikosov:2002jr}
  A.~A.~Abrikosov,
  ``Dirac operator on the Riemann sphere,''
  arXiv:hep-th/0212134.


\bibitem{Ortin:2002qb}
  T.~Ortin,
  ``A Note on Lie-Lorentz derivatives,''
  Class.\ Quant.\ Grav.\  {\bf 19}, L143-L150 (2002).
  [hep-th/0206159].













\bibitem{Bagger:2007jr}
  J.~Bagger and N.~Lambert,
  Phys.\ Rev.\  D {\bf 77}, 065008 (2008)
  [arXiv:0711.0955 [hep-th]].





\bibitem{Hosomichi:2008qk}
  K.~Hosomichi, K.~M.~Lee and S.~Lee,
  Phys.\ Rev.\  D {\bf 78}, 066015 (2008)
  [arXiv:0804.2519 [hep-th]].

\bibitem{Gomis:2008cv}
  J.~Gomis, A.~J.~Salim and F.~Passerini,
  JHEP {\bf 0808}, 002 (2008)
  [arXiv:0804.2186 [hep-th]].


\bibitem{Myers:1999ps}
  R.~C.~Myers,
  ``Dielectric branes,''
  JHEP {\bf 9912}, 022 (1999)
  [arXiv:hep-th/9910053].
  

\bibitem{Witten:1998uka}
  E.~Witten,
  ``Theta dependence in the large N limit of four-dimensional gauge theories,''
  Phys.\ Rev.\ Lett.\  {\bf 81}, 2862 (1998)
  [arXiv:hep-th/9807109].

\bibitem{Henningson:2004dh}
  M.~Henningson,
  ``Self-dual strings in six dimensions: Anomalies, the ADE-classification, and
  the world-sheet WZW-model,''
  Commun.\ Math.\ Phys.\  {\bf 257}, 291 (2005)
  [arXiv:hep-th/0405056].

\bibitem{Festuccia:2011ws}
  G.~Festuccia and N.~Seiberg,
  ``Rigid Supersymmetric Theories in Curved Superspace,''
  JHEP {\bf 1106}, 114 (2011)
  [arXiv:1105.0689 [hep-th]].


\bibitem{Pasti:2011zz}
  P.~Pasti, I.~Samsonov, D.~Sorokin and M.~Tonin,
  ``BLG and M5,''
  Phys.\ Part.\ Nucl.\ Lett.\  {\bf 8}, 209 (2011).

\bibitem{arXiv:0805.2898} 
  P.~-M.~Ho, Y.~Imamura, Y.~Matsuo and S.~Shiba,
  ``M5-brane in three-form flux and multiple M2-branes,''  JHEP\ {\bf 0808}, 014  (2008)  [arXiv:0805.2898 [hep-th]].  

\bibitem{arXiv:0806.4044} 
  K.~Furuuchi, S.~-Y.~D.~Shih and T.~Takimi,
  ``M-Theory Superalgebra From Multiple Membranes,''  JHEP\ {\bf 0808}, 072  (2008)  [arXiv:0806.4044 [hep-th]].  


\bibitem{arXiv:1006.5291} 
  C.~-H.~Chen, K.~Furuuchi, P.~-M.~Ho and T.~Takimi,
  ``More on the Nambu-Poisson M5-brane Theory: Scaling limit, background independence and an all order solution to the Seiberg-Witten map,''  JHEP\ {\bf 1010}, 100  (2010)  [arXiv:1006.5291 [hep-th]].  


\bibitem{arXiv:1008.0902} 
  A.~Gustavsson,
  ``An Associative star-three-product and applications to M two/M five-brane theory,''  JHEP\ {\bf 1011}, 043  (2010)  [arXiv:1008.0902 [hep-th]].  

\bibitem{arXiv:1012.2707} 
  A.~M.~Low,
  ``Aspects of Supersymmetry in Multiple Membrane Theories,''  arXiv:1012.2707 [hep-th].  

\bibitem{Linander:2011jy} 
  H.~Linander and F.~Ohlsson,
  ``(2,0) theory on circle fibrations,''  arXiv:1111.6045 [hep-th].  


\end{thebibliography}
\end{document}